\documentclass[fleqn,10pt]{wlscirep}
\usepackage[utf8]{inputenc}
\usepackage[T1]{fontenc}
\usepackage{txfonts}
\usepackage{amsfonts}
\usepackage{amsmath}
\usepackage{graphicx}
\usepackage{dcolumn}
\usepackage{bm}
\usepackage{overpic}
\usepackage{multirow}

\title{Geometry-controlled phase transition in vibrated granular media}

\author[1,2]{René Zuñiga}
\author[2]{Germán Varas}
\author[1,*]{Stéphane Job}
\affil[1]{Laboratoire Quartz, EA-7393, ISAE-Supm{\'e}ca, 3 rue Fernand Hainaut 93400 Saint-Ouen-sur-Seine, France.}
\affil[2]{Instituto de F\'isica, Pontificia Universidad Cat\'olica de Valpara\'iso, Avenida Brasil 2950, Valpara\'iso, Chile.}
\affil[*]{Corresponding author: stephane.job@isae-supmeca.fr}

\begin{abstract}
We report experiments on the dynamics of vibrated particles constrained in a two-dimensional vertical container, motivated by the following question: how to get the most out of a given external vibration to maximize internal disorder (e.g. to blend particles) and agitation (e.g. to absorb vibrations)? Granular media are analogs to classical thermodynamic systems, where the injection of energy can be achieved by shaking them: fluidization arises by tuning either the amplitude or the frequency of the oscillations. Alternatively, we explore what happens when another feature, the container geometry, is modified while keeping constant the energy injection. Our method consists in modifying the container base into a V-shape to break the symmetries of the inner particulate arrangement. The lattice contains a compact hexagonal solid-like crystalline phase coexisting with a loose amorphous fluid-like phase, at any thermal agitation. We show that both the solid-to-fluid volume fraction and the granular temperature depend not only on the external vibration but also on the number of topological defects triggered by the asymmetry of the container. The former relies on the statistics of the energy fluctuations and the latter is consistent with a two-dimensional melting transition described by the KTHNY theory.
\end{abstract}

\begin{document}
\flushbottom
\maketitle
\thispagestyle{empty}


\section*{Introduction}

\begin{figure}[t]
\centering
\includegraphics[width=0.95\textwidth]{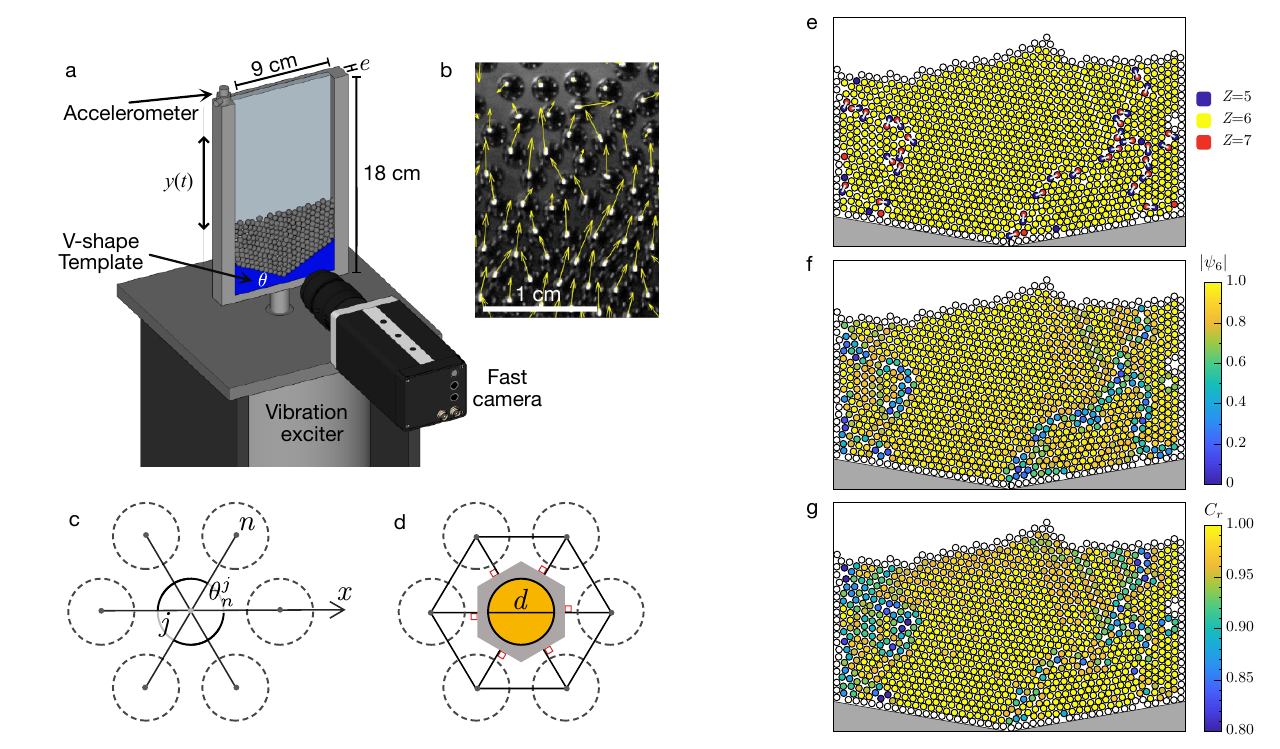}
\caption{\label{fig:setup_static}(a) Sketch of the experimental setup. (b) Image crop displaying few particles and the instantaneous velocity field (yellow arrows). (c) The local order parameter $\psi_6(j)$ probes the internal orientations of the nearest neighbors (NN) of a particle $j$, thanks to the Delaunay triangulation (DT) of the lattice. (d) The local compaction $C(j)$ is the ratio of particle cross-section (in orange) to cell area (in gray) of the Voronoi tessellation (VT, dual of the DT). Maps in a sample at rest $(\theta,\Gamma)=(10^{\circ},0)$ of (e) the coordination number $Z(j)$ (i.e. the number of NN of $j$) revealing the disclinations (isolated particles with $Z=5$ or $Z=7$) and the dislocations (pairs uniting $Z=5$ and $Z=7$, see the white lines), (f) the magnitude of the order parameter $|\psi_6(j)|$ and (g) the relative compaction $C_r(j)=C(j)/C_{hcp}$.}
\end{figure}

Driven granular matter is a classic out-of-equilibrium system exhibiting pattern-forming instabilities~\cite{Aranson2006}. These media are usually composed of many macroscopic particles that interact through short-range repulsive interactions~\cite{jaeger1996,deGennes1999}. Among various patterns, one of the most studied in the last two decades has been the solid-liquid-like phase transition, depending mainly on the driving acceleration and the packing fraction~\cite{Olafsen1998,Quinn2000,Falcon2001,Olafsen2005,Daniels2006,Reis2006,Roeller2011,Heckel2015,Clewett2016,Clewett2012,Noirhomme2021}.
In particular, granular matter exhibits patterns and instabilities that resemble those of molecular fluids~\cite{Melo1995,Umbanhowar1996} and has the ability to organize similarly to the phases of condensed matter.
The interplay between defects and vibrations in granular media is a central question: it constitutes a basic mechanism of transition to spatio-temporal disorder~\cite{Douady1989} in grains driven far from equilibrium. In highly ordered lattices, e.g. granular crystals, such an interplay is a way to trigger nonlinear instabilities leading to strong energy localization~\cite{Theocharis2009,Job2009} and spontaneous symmetry breaking~\cite{Boechler2010,Ponson2010}.
Indeed, the spatial arrangements of grains range from (i) crystalline solids, in which atoms form a perfectly periodic lattice extending in all directions, to (ii) amorphous matter, such as fluids or glasses, in which the atoms are fully disordered. The former possesses a long-range order, and the latter is both orientationally and positionally isotropic~\cite{Briand2016,Klamser2018,Rietz2018}. Specifically, in two-dimensional systems of particles, an intermediate state of matter is also possible, namely a hexatic phase. In this configuration, the atoms are distributed randomly, as in a fluid or glass so that the translational order becomes short-range, but they keep the quasi-long-range orientational order found in the crystal. The nature of this transition has been extensively studied since the 70's and predicted by Kosterlitz-Thouless-Halperin-Nelson-Young (KTHNY) theory~\cite{Strandburg1988,Kosterlitz2017}. Here, the solid-liquid melting phase is mediated by topological defects, such as vortices in a superfluid or dislocations and disclinations in a crystal~\cite{Kosterlitz1972,Kosterlitz1973,Berthier2011,Chen2021}. This concept has recently been taken to granular media~\cite{Fortini2015,Sun2016,Ramming2017,Thorneywork2017,Cao2018,Arjun2020} and ensembles of hard discs~\cite{Binder2002,Tuckman2020}.
Unraveling the nature of the phase transition in 2D vibrated granular media however remains an open and intense debate~\cite{Prevost2004,Goetzendorfer2005,Melby2005,Pastor2007,Bernard2011,Castillo2012,Qi2014,Kapfer2015,Mujica2016}, inquiring as to whether it is a first-order one or a continuous one~\cite{Strandburg1988}. As a matter of fact, transitions triggered by the unbinding of dislocations, as predicted by the KTHNY scenario, are likely of the second-order continuous type~\cite{Kapfer2015} whereas the observation of phase coexistence is a fingerprint of a first-order transition~\cite{Bernard2011}. For instance, the coexistence of liquid-solid states has been observed in a quasi-one-dimensional driven granular media, where a first-order like transition was found to be mediated by waves and triggered by negative compressibility, as in a van der Waals gas model~\cite{Clerc2008}.
More recently, the role of the energy dissipation~\cite{Komatsu2015} and of the container's roughness~\cite{Downs2021} have been investigated, demonstrating that particles inelasticity~\cite{Komatsu2015} and surface topography~\cite{Downs2021} both have the ability to alter the nature and the order of the solid-liquid transition, from a two-step continuous one (involving an intermediate hexatic phase) to a discontinuous first-order-like one (involving an intermediate state where liquid and solid coexist).
Practically, understanding how to control the amount of disorder and agitation of particles is meaningful in contexts relevant to industrial processing of granular materials, like segregation and mixing of seeds and pills~\cite{Ottino2000,Barker2021} or vibrations mitigation~\cite{Saluena1999,Sanchez2012,Sack2013,Masmoudi2016,Ferreyra2021}, for instance. Here, our objective is twofold. On the one hand, we aim at quantifying how the geometry of the container affects the temperature (i.e. the internal energy) of the system at constant energy injection. On the other hand, we explore how this modification impacts the amount of disorder within the system: how the system transits from a solid-like to fluid-like state for different container shapes. In particular, we seek to unravel whether one can tune the state of the matter to a given solid-to-fluid volume fraction by maintaining a constant energy injection into the system.
The experimental setup under consideration is shown in Fig.~\ref{fig:setup_static}(a). It consists of a transparent vertical Hele-Shaw cell filled with monodisperse spherical beads. The bottom of the cell is a V-shape wall with a given angle $\theta$. The cell is vibrated vertically by a shaker with displacement amplitude $A$ and frequency $f=\omega/2\pi=30$~Hz, such that the dimensionless acceleration is $\Gamma=A\omega^2/g$, being $g$ the gravitational acceleration. A fast camera placed in front of the sample records a picture (at rest) or a movie (under vibrations) from which one extracts, firstly, the instantaneous position, displacement and velocity fields of every particle in the cell, and secondly, all the topological features of the lattices (see Methods).


\section*{Results and discussion}

We describe the topology of a two-dimensional lattice of particles (in terms of defects, local compaction and order parameter) as a function of the container shape, first at rest and then as a function of the amplitude of vibrations. A phase transition is revealed by tracking how the solid-to-fluid volume fraction evolves as a function of both the injected energy and the container's geometry. We also perform a statistical analysis of the velocity field to evaluate how the thermal fluctuations are affected within the same parameters space. We show that a Maxwell-Boltzmann description of the vibrations ties all these features together to estimate the energy to initiate a topological defect.


\subsection*{\label{sec:rest}Disorder and defects at rest}

First, we analyze the system at rest, under a quasi-static rain-like deposit of the grains, in order to exemplify and study its spontaneous ordering. The coordination number $Z$, the relative compaction $C_r$ and the order parameter $|\psi_6|$ (see the definitions in Methods) are represented in Fig.~\ref{fig:setup_static} for a V-shape bottom surface with $\theta=10^{\circ}$. We observe that a large number of grains settle in a dense hexagonal arrangement, with $Z=6$, $C_r\simeq1$ and $|\psi_6|\simeq1$ locally. Topological defects originating from the three edges of the container's base initiate quasi-one-dimensional fractures extending throughout the lattice, until the top surface. These long-range defects accommodate the symmetry breaking of the V-shape base, by impeding the tendency to form a hexagonal lattice. As shown in Fig.~\ref{fig:setup_static}(f,g) these branches are both loose ($C_r<1$) and less ordered ($|\psi_6|<1$) defects. They correspond to weak and brittle unconsolidated regions, where the particles are free to move or slide one on the others, i.e. where melting may initiate more favorably under vibrations. The potential energy stored in these defects is taken from the energy implemented to prepare the lattice, which may likely contribute later, either to increase the temperature of agitation or to lower the energy required to transit from a weakened solid phase to a fluid phase.


\subsection*{\label{sec:dynamics}Disorder and defects under vibrations}


\begin{figure}[t]
\centering
\includegraphics[width=0.95\textwidth]{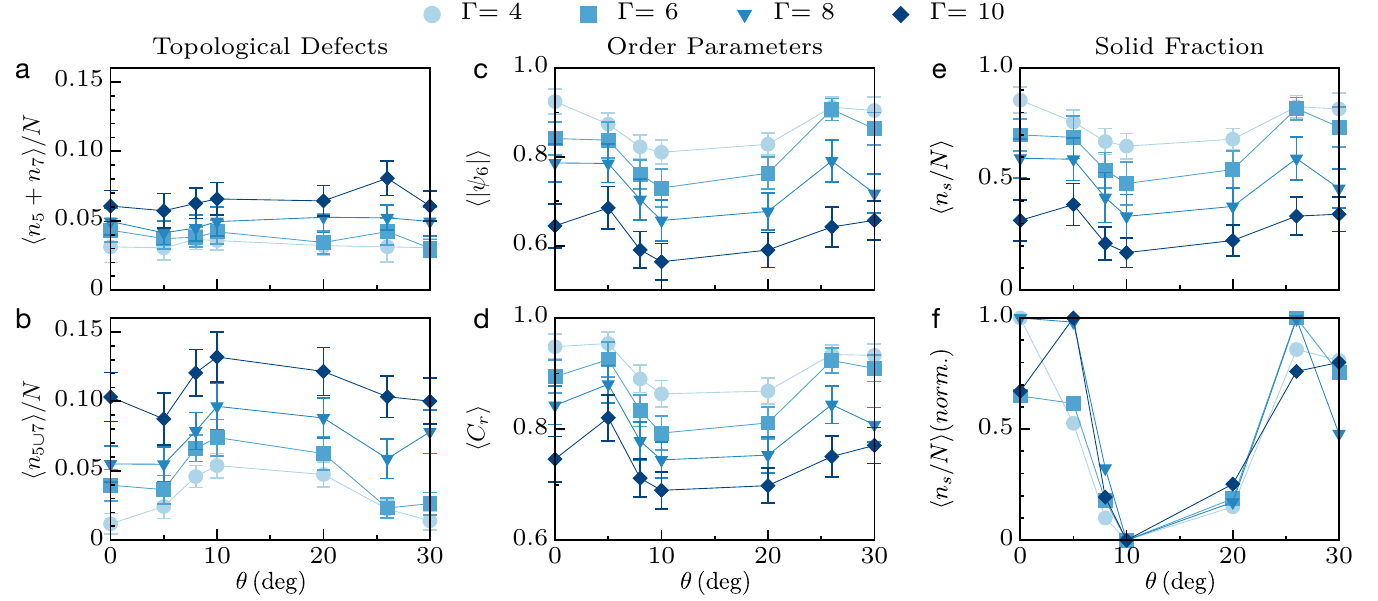}
\caption{\label{fig:dynamics} (a) Fraction of disclinations, (b) fraction of dislocations, (c) order parameter, (d) relative compaction, (e) solid fraction and (f) normalized solid fraction as a function of V-shape angle $\theta$ and acceleration amplitude $\Gamma$. (a,b,e,f) are averages over time whereas (c,d) are averages over time and particles.}
\end{figure}

We repeat the estimations of $Z$, $C_r$, $|\psi_6|$ and defect number versus time and space under dynamic conditions, for four acceleration amplitudes $\Gamma$ (from $4$ to $10$, increasing $A$ at constant $f$) and seven V-shape angles $\theta$. A V-shape with $\theta=0^{\circ}$ or $\theta=30^{\circ}$ matches the symmetry of a hexagonal lattice, whereas it breaks this symmetry for angles in between. First, we checked that the temporal evolution of $Z$ is relatively constant in time, with small fluctuations, revealing around six neighbors on average (see Fig.~S3 in the Supplementary Information). This indicates that the system spontaneously maintains a hexagonal arrangement independent of the angle of the base and the energy of vibration. Looking in detail (see for instance the animations, Fig.~S6 in the Supplementary Information) reveals that, as for the static case shown in Fig.~\ref{fig:setup_static}(e), a small fraction of particles can temporarily lose ($Z=5$) or gain ($Z=7$) one nearest neighbor (NN). These 5-folded and 7-folded defects are referred to \emph{disclinations} when isolated: they break the local orientational symmetry in a crystal~\cite{Hattar2011}. In a perfect hexagonal lattice, they originate in connected pairs (named $5\cup7$ in the following) referred to \emph{dislocations}, that break the translational order of the lattice~\cite{Sun2016}.
Dynamically, the creation-annihilation and the mobility of topological defects (both observed in our system, see Fig.~S2 in the Supplementary Information) are the basic relaxation mechanisms by which a lattice accommodates disorder~\cite{Shen2019}.
According to KTHNY theory, dislocations appear at the solid-to-hexatic phase transition whereas disclinations, resulting from the unbinding of dislocations at a higher temperature, appear at the hexatic-to-fluid transition~\cite{Komatsu2015}. In Fig.~\ref{fig:dynamics}(a,b), $\langle n_5+n_7\rangle$ and $\langle n_{5\cup7}\rangle$ denote the time-averaged number of disclinations and dislocations, respectively; the complementary fraction, at least $80\%$ of the sample, corresponds to a crystal state, $Z=6$.
According to expectations, the number of topological defects increases with the amplitude of the vibration $\Gamma$. In particular, the presence of disclinations and dislocations reveals that the system has undergone the two-phases transitions predicted by the KTHNY theory, with part of the system being in a fluid-like state (i.e. amorphous) and the other being in a crystal-like state (i.e. solid or hexatic). A pure hexatic phase would contain dislocations only. Remarkably, the shape of the container also affects the fraction of both topological defects at constant energy injection: they are maximized in a container that breaks the symmetry of the sample, see Fig.~\ref{fig:dynamics}(a,b) at $\theta=10^{\circ}$. These observations are consistent with the evolution of the time-and-space averaged $\langle C_r\rangle$ and $\langle |\psi_6|\rangle$, see Fig.~\ref{fig:dynamics}(c,d): increasing the injected energy or breaking the symmetry of the lattice induces a more diluted and disordered structure, as one can expect when a crystal melts. Interestingly, both $C_r$ and $|\psi_6|$ present a minimum close to $\theta=10^{\circ}$, for which the number of topological defects is also maximum. Finally, note how correlated are the curves of $C_r$ and $|\psi_6|$, revealing a linear proportionality between order and density (see also Fig.~S4 in the Supplementary Information).


\begin{figure}[t]
\centering
\includegraphics[width=0.95\textwidth]{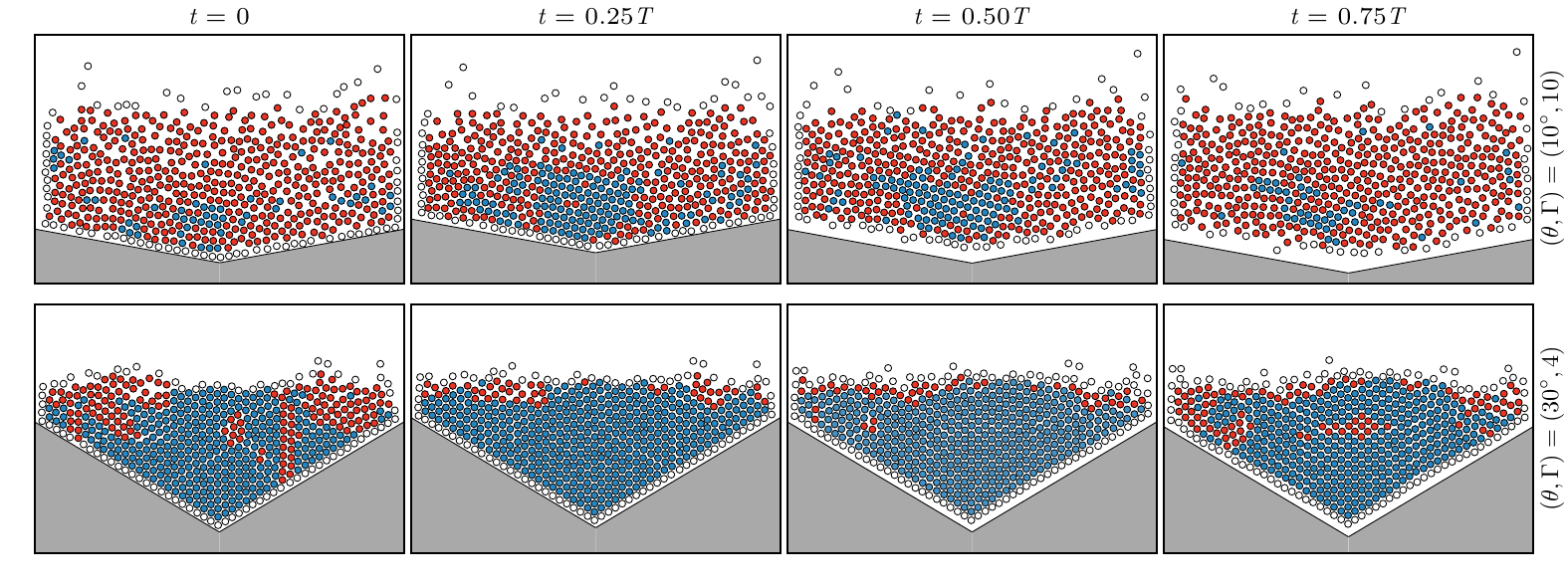}
\caption{\label{fig:movie}Snapshots of solid-like (blue) and fluid-like (red) particles at four instants of vibrations within a period of oscillation. Top: $(\theta,\Gamma)=(10^{\circ},10)$. Bottom: $(\theta,\Gamma)=(30^{\circ},4)$. The particles located at the outer edge of the ensemble, shown in white, are not used in the analysis.}
\end{figure}


\subsection*{\label{sec:transition}Phase detection and evolution}

Locally, a cluster of particles can be considered as a solid (jammed, dense, crystallized) or a fluid (unjammed, loose, amorphous) depending on criteria on the order parameter~\cite{Olafsen2005,Reis2006,Briand2016,Klamser2018,Berthier2011,Chen2021,Sun2016,Ramming2017,Arjun2020,Binder2002,Melby2005,Bernard2011,Castillo2012,Qi2014,Mujica2016,Komatsu2015,Downs2021,Luu2013,Han2008,Buttinoni2017,Kansal2002,Chakrabarti1998} and/or on the packing fraction~\cite{Quinn2000,Falcon2001,Reis2006,Roeller2011,Heckel2015,Kapfer2015,Komatsu2015,Clewett2016,Sun2016,Thorneywork2017,Rietz2018}. Based on our observations (see Fig.~S4 in the Supplementary Information), inspecting the probability density function of the local order parameter in lattices at several driving amplitude and containers shape, reveals a sharp peak centered on $|\psi_6|\simeq1$ (i.e. hexagonal crystal chunks) and a wider one spanning around $|\psi_6|\simeq0.5$ (i.e. random amorphous phases). These two contributions are well separated by a local minimum located near $|\psi_6^\ast|=0.85$; this value constitutes a good candidate for a criterion to distinguish solid-like and fluid-like phases. Alternatively, we checked that a criterion on the local compaction, $C_r^\ast=0.9$, provides a similar discrimination level owing to the linear correlation between $|\psi_6|$ and $C_r$ (see Fig.~S4 in the Supplementary Information). We thus differentiate phases at the particle level depending on the local order parameter, whether the particle $j$ and its NN are in
\begin{equation}
\left \{ \begin{array}{r}
\mbox{a crystallized and dense solid-like state if }|\psi_6(j)|\geq0.85\mbox{,} \\
\mbox{an amorphous and loose fluid-like state if }|\psi_6(j)|<0.85\mbox{.} \end{array} \right.\label{eq:threshold}
\end{equation}
This definition is used in Fig.~\ref{fig:dynamics}(e,f) to represent the time-averaged solid fraction $\langle n_{S}/N\rangle$ in a vibrated lattice, as a function of $\Gamma$ and $\theta$, where $n_S(t)$ is the instantaneous number of solid-like particles according to Eq.~\ref{eq:threshold} and $N$ is the total number of particles in the sample. The solid fraction $n_S(t)/N$ fluctuates at the frequency of the external driving (see the Fig.~S4 in the Supplementary Information) and the time-average quantifies how the coexistence of phases evolves in the system: in the most fluidized case, one finds about $5\%$ of particles in a solid-like state, whereas, for the least fluidized case, there is approximately $60\%$ of such particles. As expected, the largest solid fractions are obtained for the smallest acceleration. In contrast, at constant acceleration, a minimum of the solid fraction is reached when the container and the lattice are geometrically dissimilar. The maximum is reached when they have the same symmetries, at $\theta=0^{\circ}$ and $\theta=30^{\circ}$. More precisely, a min-max normalization of the curves presented in Fig.~\ref{fig:dynamics}(e) shows that the geometry at which the minimum solid fraction is reached, $\theta=10^{\circ}$, does not depend on the driving amplitude, see Fig.~\ref{fig:dynamics}(f). The fact that the normalized data collapse on a master curve demonstrates that the solid fraction can be tuned independently by acting on the geometry of the container only, at constant input energy. Consistently, the solid fraction is minimum at $\theta=10^{\circ}$ because at this value, the sample contains the largest fraction of topological defects in addition to being in the loosest and the more disordered state, see Fig.~\ref{fig:dynamics}(a-d).
The condition given in Eq.~\ref{eq:threshold} also facilitates the instantaneous monitoring of the phases, as illustrated in Fig.~\ref{fig:movie} for two contrasted sets of parameters $(\theta,\Gamma)=(30^{\circ},4)$ and $(\theta,\Gamma)=(10^{\circ},10)$. According to Fig.~\ref{fig:dynamics}, the former case corresponds to a geometry matching the symmetry of a hexagonal lattice considered at low driving amplitude whereas the latter case boosts the solid-to-liquid phase ratio, in terms of both the container shape and the energy injection. Each column display instantaneous snapshots of the system at four different stages of oscillations, $t/T=[0, 1/4, 1/2, 3/4]$. Blue stands for a solid-like particle and red stands for a fluid-like particle. For the asymmetric configuration at large driving amplitude, $(\theta,\Gamma)=(10^{\circ},10)$, the displacement field appears disorganized and diluted: the sample is more fluidized. In contrast, the symmetric configuration within the smallest driving amplitude, $(\theta,\Gamma)=(30^{\circ},4)$ appears more organized and denser, with wide clusters of crystallized and jammed structures. Both examples demonstrate the coexistence of solid-like and fluid-like phases in the sample, on average. Remarkably, the value of the solid fraction is transient as it oscillates around an average value at the driving frequency (see Fig.~S4 in the Supplementary Information), unlike the horizontal granular beds vibrated out-off-plane, in which the phases are quasi-stationary~\cite{Prevost2004,Clerc2008}.


\subsection*{Statistical analysis of the thermal fluctuations}

Analyzing now the fluctuations of the velocity field, obtained by tracking the trajectories of all particles (see Methods), provides information (i) on how the kinetic energy is statistically distributed inside the granular medium and (ii) on how the transfer of external vibration into internal agitation is affected by the geometry and the driving. The velocity fluctuations of the particle $j$ in the $x$ (horizontal) or $y$ (vertical) directions is $\tilde{v}_{x,y}(j,t)=v_{x,y}(j,t)-\bar{v}_{x,y}(t)$. The quantity $\bar{v}$ denotes the ensemble average, over all particles at a given instant: it thus corresponds to the in-plane component of the instantaneous velocity of the center of mass of the monodisperse sample, $\bar{v}_{x,y}(t)=v_{x,y}^{CM}(t)$ (see Fig.~S5 in the Supplementary Information). In the following, we analyze qualitatively how is distributed the energy of agitation associated with these fluctuations, before quantifying them more systematically owing to the analysis of their first four statistical moments as a function of $\theta$ and $\Gamma$. In particular, the inspection will show that the thermal agitation is noticeably anisotropic but tends to be satisfactorily approximated by the Maxwell-Boltzmann distribution at large amplitude and asymmetry.


\begin{figure}[t]
\centering
\includegraphics[width=0.95\textwidth]{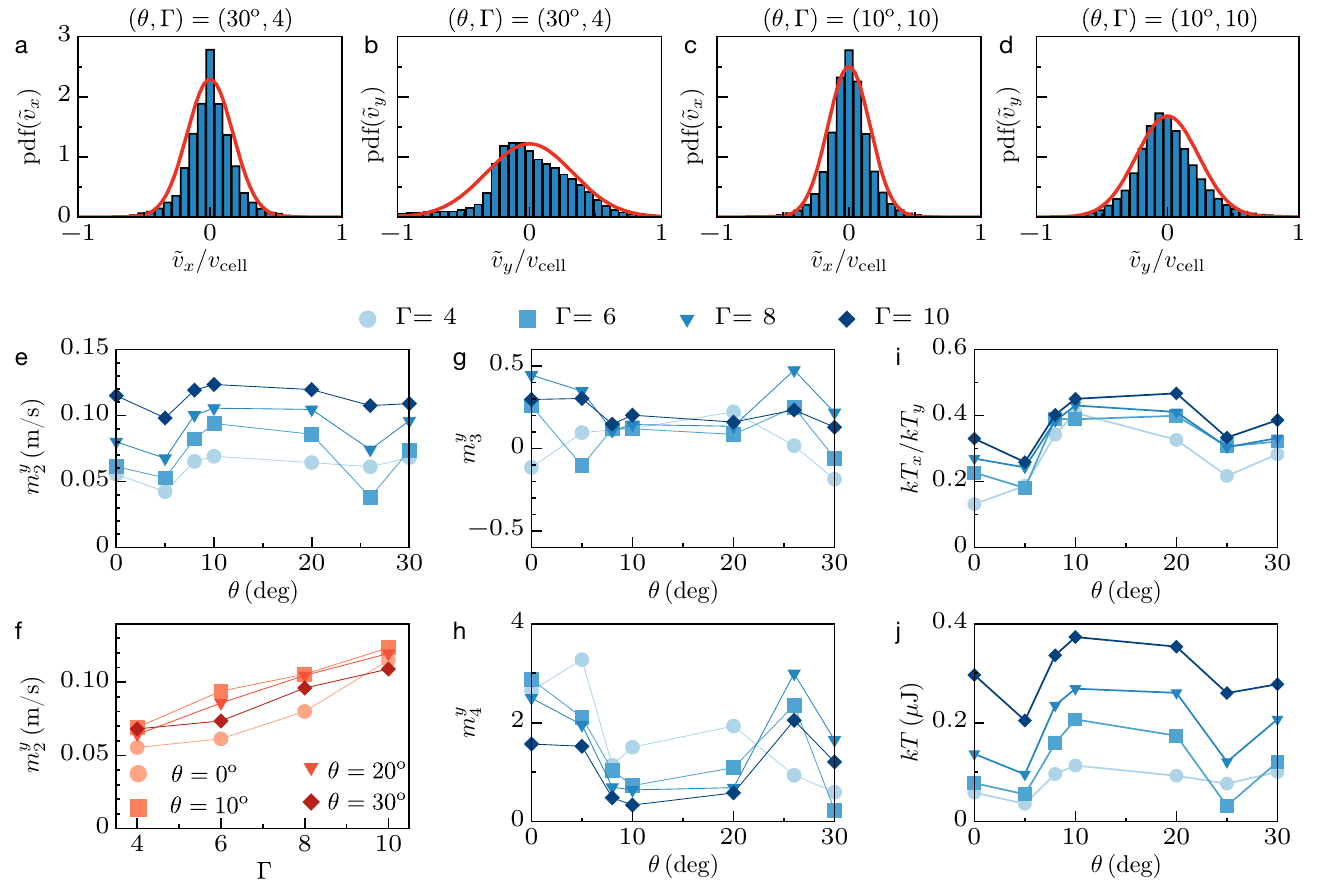}
\caption{\label{fig:statistics}Probability density function of the horizontal and vertical velocity fluctuations, $\tilde{v}_{x,y}/v_{\text{cell}}$, for (a,b) $(\theta,\Gamma)=(30^{\circ},4)$ and (c,d) $(\theta,\Gamma)=(10^{\circ},10)$. The red lines in (a-d) are the best fits with the normal distribution, see Eq.~\ref{eq:normal_distrib}. Vertical standard deviation $m_2^y$ versus (e) $\theta$ and (f) $\Gamma$. (g) Vertical skewness $m_3^y$ and (h) vertical kurtosis $m_4^y$ versus $\theta$ for different $\Gamma$. (i) Horizontal-to-vertical temperature ratio, $kT_x/kT_y$, and (j) mean temperature, $kT=(kT_x+kT_y)/2$, versus $\theta$ for different $\Gamma$.}
\end{figure}


\textit{Probability density function \---}
In Fig.~\ref{fig:statistics}(a-d), we present two examples of probability density function (PDF) of the velocity fluctuations in the $(x,y)$ plane, at low and high fluidizations (same cases as in Fig.~\ref{fig:movie}). In each of the four plots, the red curve stands for the best adjustment with the normal distribution,
\begin{equation}
\mbox{pdf}(\tilde{v}_{x,y})=\sqrt{m/2\pi kT_{x,y}}\exp{(-m\tilde{v}_{x,y}^2/2kT_{x,y})}
\mbox{ with }
m_2^{x,y}=\mbox{std}(\tilde{v}_{x,y})=\sqrt{kT_{x,y}/m},\label{eq:normal_distrib}
\end{equation}
where the standard deviation $m_2^{x,y}$ is estimated over all particles and times; $m_2^{x,y}$ relates to the granular temperature~\cite{Olafsen1998,Olafsen1999,Losert1999,Losert2000,Baxter2006} in the horizontal or vertical directions, $kT_{x,y}$ given in Joules, being $m$ the mass of a particle and $k$ a macroscopic equivalent of the Boltzmann's constant.
A satisfactory fit of the PDF of the individual components with a Gaussian distribution means that the magnitude of the fluctuations velocity vector follows the Maxwell-Boltzmann distribution~\cite{Barrat2002,Meer2006} (see below), though a careful inspection allows to reveal slight discrepancies. One sees for instance that both PDF in the vertical direction shown in Fig.~\ref{fig:statistics}(b,d) are slightly skewed on the left, indicating an excess (resp. a lack) of negative (resp. positive) velocities. This stems from the asymmetry introduced by the gravity field (see the motions in Fig.~S5 in the Supplementary Information), in addition to the inelastic rebound at the collision between the free-falling granular lattice and the bottom of the container~\cite{Ferreyra2021}. Consistently, we observe that unlike the vertical velocity distribution, the distribution of the horizontal velocity fluctuations is better centered in both examples.
Also noticeable, the horizontal and the vertical PDFs appear sharper than the normal distribution. Such a non-Gaussian trend is also consistent with the dissipative nature of the interactions between particles~\cite{Noije1998,Losert1999,Saluena1999}, leading to an excess of those with low velocity/low kinetic energy. However, the sharpening of the PDFs has been shown to have a negligible effect on the estimation of the temperature in systems driven stationarily~\cite{Noije1998}.
Finally, we note that the velocity fluctuations are significantly anisotropic: the width of the PDFs in the horizontal direction is thinner than in the vertical direction, compare for instance Figs.~\ref{fig:statistics}(a,c) and~\ref{fig:statistics}(b,d). Such anisotropy is a well-known consequence of the unidirectional driving~\cite{Barrat2002,Meer2006}, the energy being injected in the vertical direction.


\textit{Statistical moments \---}
We rationalize now the inspection of the statistical properties by estimating systematically the $n$-th central moments $\mu_n^{x,y}=\langle(\tilde{v}_{x,y})^n\rangle$ of the horizontal/vertical agitation versus $\Gamma$ and $\theta$. Here, the average, $\langle\dots\rangle$, is performed over all particles and times.
More details on the dynamical evolution of the statistical moments are provided in the Supplementary Information, see for instance the time-dependent boxplots shown in Fig.~S5.
On average, the first moment is exactly zero, $\mu_1=0$, since the instantaneous velocity of the center of mass of the granular slab is removed from all the particles in each frame. The second moment provides the standard deviation, $m_2=\sqrt{\mu_2}$, which is a measure of the magnitude of the fluctuations. The skewness $m_3=\mu_3/\mu_2^{3/2}$ indicates the asymmetry of the PDF and the normalized kurtosis $m_4=\mu_4/\mu_2^2-3$ quantifies its sharpness~\cite{Baxter2006}. It is worth noting that all the moments are calculated using unbiased estimators. Also, the high order statistical moments being more sensitive to the experimental noise than the lower ones, we improve the accuracy of both $m_3$ and $m_4$ by discarding irrelevant outliers data (defined as the velocity fluctuations whose magnitude is larger than three times the standard deviation). As a matter of fact, the skewness and the kurtosis are zero for a normal distribution, $m_3=m_4=0$: both these statistical moments thus measure the relative degree of similarities between the observed fluctuations and the Maxwell-Boltzmann distribution~\cite{Baxter2006}. The evolution of the different moments as a function of $\Gamma$ and $\theta$ are summarized in Fig.~\ref{fig:statistics}(e-h). First, we observe that the standard deviation increases monotonically with the magnitude of the excitation $\Gamma$, see Fig.~\ref{fig:statistics}(e,f), revealing an expected augmentation of the internal thermal agitation with the external driving amplitude. Interestingly, $m_2^y$ also depends on $\theta$ at given $\Gamma=\mbox{const}$. In particular, the magnitude of the fluctuations is boosted at intermediate angles ($10^{\circ}<\theta<20^{\circ}$), where one previously found that the amount of topological defects is maximum and the order parameter, the compaction and the solid fraction are minimum, see Fig.~\ref{fig:dynamics}. More asymmetry thus induces more defects, disorder and looseness, leading to stronger agitation.
The evolution of the skewness in the vertical direction is also consistent with the examples given in Fig.~\ref{fig:statistics}(a-d). It is slightly positive but very close to a normal distribution ($m_3^y\simeq0.1$) for intermediate angle ($10^{\circ}\leq\theta\leq20^{\circ}$). For the more symmetric configurations ($\theta\simeq0^{\circ}$ and $\theta\simeq30^{\circ}$) a larger variability of $m_3^y$ is observed. However, these deviations do not reveal a clear and monotonic tendency, such that on average, the skewness could be considered roughly independent on $\Gamma$ and $\theta$.
In contrast, the normalized kurtosis of the vertical component of the velocity appears more significantly affected by the driving amplitude and the geometry. In particular, it tends to that of a normal distribution, $m_4^y\simeq0$, for intermediate angles ($10^{\circ}<\theta<20^{\circ}$). In this region, the larger the driving amplitude, the closer to a normal distribution. Unlike the skewness, the normalized kurtosis monotonically rises when the geometry becomes more symmetric ($\theta=0^{\circ}$ and $\theta=30^{\circ}$), indicating a deviation from a Gaussian distribution resulting from the container geometry. Here, note that the contrast between symmetric and asymmetric geometries tends to fade as the amplitude of the driving increases.
Nevertheless, the ratio of horizontal-to-vertical standard deviations confirms the systematic anisotropy of the temperature field~\cite{Meer2006,Barrat2002}, see Fig.~\ref{fig:statistics}(i). Interestingly enough, the ratio appears relatively $\Gamma$-independent but $\theta$-dependent: the anisotropy is more pronounced for symmetric containers than for asymmetric ones.
Further quantitative analysis (see below) of the thermal agitation will be carried on with the definition of an average isotropic-like temperature, $T=(T_x+T_y)/2$, as proposed by Barrat et al.~\cite{Barrat2002}. The plot of such a temperature, shown in Fig.~\ref{fig:statistics}(j), confirms that for a given container's geometry, the injected energy is converted into thermal agitation that increases monotonically with the input. It also demonstrates that, unlike classical fluids or solids, the internal temperature can be changed significantly by acting on the container's shape only, at constant external excitation; in particular, the most asymmetrical cases boost the temperature by approximately a factor two.
Here, the V-shapes with intermediate angles enhance both the magnitude of the fluctuations and the temperature isotropy, by scattering the vertically injected momentum more efficiently, out of the gravity and out of the lattice's symmetry axes.
This efficiency relies on particles' mobility, i.e. on topological defects: a defect is activated when the displacement of a particle is energetically possible, see the Sec.~6 in the Supplementary Information, without preferred direction. Although the magnitude and the isotropy of the temperature appear related, this suggests a leading effect of the former on the generation of disorder: the larger the temperature, the more topological defects, see the correlation between Figs.~\ref{fig:dynamics}(a,b) and Fig.~\ref{fig:statistics}(j), whereas the temperature anisotropy is only a function of $\theta$, see Fig.~\ref{fig:statistics}(i).

\begin{figure}[t]
\centering
\includegraphics[width=0.95\textwidth]{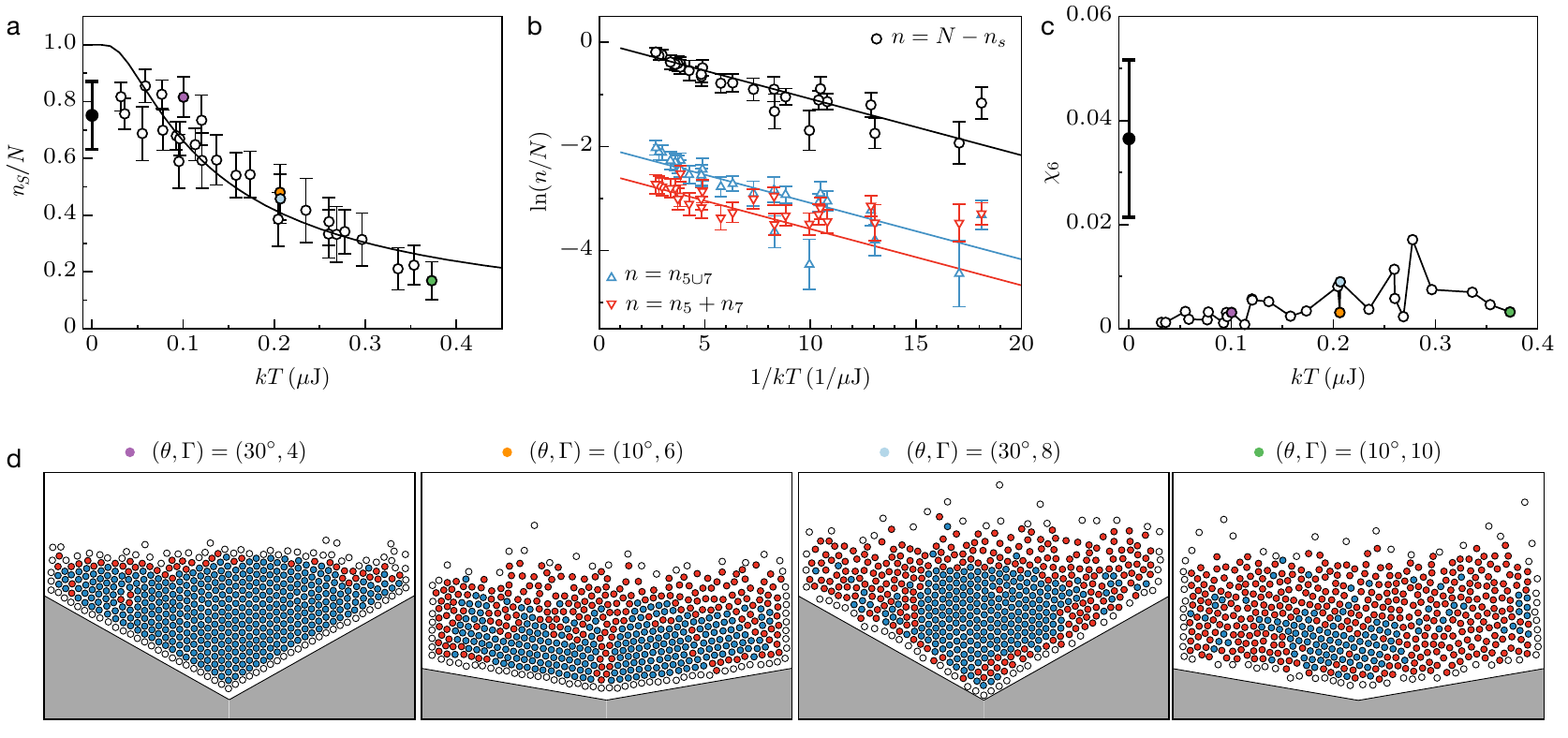}
\caption{\label{fig:merge} (a) Solid fraction, (b) fluid fraction superimposed with defects fractions and (c) susceptibility, as a function of temperature, merging all the data shown in Figs.~\ref{fig:setup_static},~\ref{fig:dynamics} and~\ref{fig:statistics} at various $(\theta,\Gamma)$ into a single data set. The black curve in (a,b) stands for the Maxwell-Boltzmann cumulative density function $\mbox{cdf}(E,T)=1-\exp{(-E/kT)}$ with $E=108$~nJ, see Eq.~\ref{eq:solid_fraction_MB}. The blue and red curves in (b) are guidelines parallel to the Maxwell-Boltzmann approximation. Black bold markers located at $T=0$ in (a) and (c) correspond to the average data obtained from ten realizations of static heaps at each angle, as seen for instance in Fig.~\ref{fig:setup_static}. (d) Snapshots of solid-like (blue) and fluid-like (red) particles, as in Fig.~\ref{fig:movie}, showing weak (first), intermediate (second and third) and large (fourth) temperatures and solid fractions; note the two intermediate cases with different $(\theta,\Gamma)$ but similar $T$ and $n_S$.}
\end{figure}

\textit{Maxwell-Boltzmann distribution \---}
The statistical analysis thus demonstrates that the normal distribution given in Eq.~\ref{eq:normal_distrib} is relevant for both components of the velocity fluctuations at all amplitudes within the intermediate range of V-shape angles, whereas it appears limited to the highest excitation magnitudes in the symmetric configurations.
Nevertheless, extending the validity of the Ansatz over the whole range of probed parameters and conceding a homogeneous isotropic temperature~\cite{Barrat2002}, entails that the magnitude of the velocity vector fluctuations, $\tilde{v}=|\vec{\tilde{v}}|$ with $\vec{\tilde{v}}=\tilde{v}_x\vec{x}+\tilde{v}_y\vec{y}$, can be approximated by the Maxwell-Boltzmann distribution~\cite{Barrat2002,Meer2006}, $\mbox{pdf}(\tilde{v})=(m\tilde{v}/kT)\exp(-m\tilde{v}^2/2kT)$.
Consequently, the Maxwell-Boltzmann cumulative density function, $\mbox{cdf}(\tilde{v})=\int_0^{\tilde{v}}{\mbox{pdf}(\tilde{v})d\tilde{v}}$, would represent the probability to find a particle with an energy in between $0$ and $E=m\tilde{v}^2/2$ in a system at temperature $T$. As a consequence, if $E$ stands for the energy threshold required to generate a dislocation, then all the particles with lower energy statistically pertain to a solid-like phase; therefore, the solid fraction would read as
\begin{equation}
n_s/N=\mbox{cdf}(E,T)=1-\exp{(-E/kT)}.\label{eq:solid_fraction_MB}
\end{equation}
Figure~\ref{fig:merge}(a) represents the solid fraction given in Fig.~\ref{fig:dynamics}(e) versus the temperature shown in Fig.~\ref{fig:statistics}(j), in addition to the average solid fraction detected in the static case ($T=0$). The latter is obtained from ten realizations of static heaps for each angle (see Methods), as seen in Fig.~\ref{fig:setup_static}(e,f,g). Fitting the Eq.~\ref{eq:solid_fraction_MB} to the dataset at $T>0$ provides a measure of the energy threshold, $E=(108\pm28)$~nJ. For the sake of comparison, this value corresponds to a fraction of the work done by the weight of a particle to elevate it by one diameter, $E\simeq(0.16\pm0.04)\times mgd$. It is coherent with the gravitational potential energy difference per particle, between an elementary hexagonal crystal standing vertically, and a 5-folded (with one void) or 7-folded (with an extra NN) topological defect. The magnitude of the fitted energy $E$ thus relies on the typical energy to create a topological defect (here a dislocation) in a hexagonal lattice.
Alternatively, in Fig.~\ref{fig:merge}(b), the same dataset is represented in a lin-log scale: the slope between the logarithm of the fluid fraction and the inverse of the temperature is $-E$. Strikingly, both fractions of defects (disclinations and dislocations) taken from Fig.~\ref{fig:dynamics}(a,b), have a trend versus the temperature which is similar to that of the fluid fraction: all three increase with $T$ at the same rate, in agreement with the assumption that $E$ in Eq.~\ref{eq:solid_fraction_MB} relies on the energy to create a topological defect. Consistently, the intercepts of the black, the blue and the red lines (i.e. the fractions at infinite temperature) in Fig.~\ref{fig:merge}(b) reveals approximately $1$ topological defect per $6$ fluid-like particles, in agreement with a trivial expectation in a hexagonal lattice.
Irrespective of Eq.~\ref{eq:solid_fraction_MB}, another important feature of the results presented in Fig.~\ref{fig:merge}(a) is the collapse of the dataset on a master curve which solely depends on the temperature, regardless of the V-shape angle. This reveals an indirect effect of the geometry on the phases under vibrations: the V-shape affects the temperature, which in turn intrinsically modifies the solid-to-fluid ratio. As an example, the two colored markers close to $kT=0.2$~$\mu$J in Fig.~\ref{fig:merge}(a) correspond to different geometries and amplitudes, $(\theta,\Gamma)=(10^{\circ},6)$ and $(\theta,\Gamma)=(30^{\circ},8)$, but lead to similar temperature and solid fraction nonetheless, as seen also in Fig.~\ref{fig:merge}(d).
In other words, this means that one can reach a given operating point, defined in terms of disorder and agitation, via any combination of the container's shape and injected energy. Without loss of generality, it is thus possible to fix the system's state at a minimal energy cost with an appropriate shape.


\subsection*{Order parameter susceptibility across phase transition.}

Finally, insight on the nature of the observed phase transition and the location of the transition points is provided here. In Thermodynamics, the susceptibility~\cite{Binder1987} provides such information, which diverges discontinuously in a first-order transition or as a power-law singularity in a second-order transition. In both case, a sharp peak of the susceptibility indicates a transition point. In particulate systems, an analog of the thermodynamic susceptibility~\cite{Weber1995} exists and has proven reliable to probe their phase transitions~\cite{Han2008,Komatsu2015,Sun2016,Downs2021}, e.g. to resolve accurately the hexatic phase in granular media~\cite{Sun2016}.
In 2D lattices, two parameters are relevant, depending on the nature of the symmetry broken across the phase transition~\cite{Han2008}, namely the translational and the orientational order parameter susceptibility, $\chi_T$ and $\chi_6$.
In the frame of a two-step transition, $\chi_T$ (revealing the solid-hexatic transition) diverges at a lower temperature than $\chi_6$ (revealing the hexatic-fluid transition), whereas both diverge at the same temperature in the case of a first-order transition~\cite{Downs2021}.
Practically, we make use of $\chi_6$ in this study, which is defined as~\cite{Weber1995}
\begin{equation}
\chi_6 = \langle|\bar{\psi}_6^2(t)|\rangle-\langle|\bar{\psi}_6(t)|\rangle^2  = \mbox{std}^2(|\bar{\psi}_6(t)|),
\label{eq:susceptibility}
\end{equation}
where $\bar{\psi}_6(t)=\sum_{j=1}^{N}{\psi_6(j,t)}/N$ is the instantaneous ensemble average and $\langle \ldots\rangle$ and $\mbox{std}(\ldots)$ denote the mean and standard deviation over time. From this definition, $\chi_6$ is also equal to the square of the standard deviation of $|\bar{\psi}_6|$. Therefore, it reveals large fluctuations of the lattice's order parameter, across a transition point~\cite{Weber1995,Han2008,Komatsu2015,Sun2016,Downs2021}.
Figure~\ref{fig:merge}(c) presents the susceptibility $\chi_6$ as a function of the temperature $T$ under vibrations, in addition to the estimation of $\chi_6$ at rest, from the repeated sample preparations (see Methods). A clear peak emerges at $T=0$ followed by a smaller plateau at any $T>0$, indicating that the loss of orientational symmetry due to the unbinding of dislocations into disclinations, occurs in the static regime.
Indeed, disclinations and dislocation are found in all static configurations, arising from the geometrical asymmetry of the container during the rain-like preparation of the samples, see Fig.~S1 in the Supplementary Information.
It results that the translational symmetry breaking, leading to the divergence of $\chi_T$ owing to the appearance of dislocations, should occur at the same temperature, $T=0$.
The latter assertion is corroborated, alternatively, by analyzing the trends of the Lindemann parameter~\cite{Zahn2000,Han2008,Sun2016} (a measure of the mean distance traveled by a particle relatively to nearest neighbors) as a function of time, see Fig.~S7 in the Supplementary Information. The Lindemann parameter does not plateau at long durations in any probed configurations, suggesting that the translational symmetry breaks~\cite{Zahn2000} at vanishing temperature, $kT=0$, concurrently with the loss of orientational order.
This means that the observed phase transition is of the first-order type, in agreement with the observation of phases coexistence, another hallmark of such a transition~\cite{Bernard2011}.
It is worth noting that our analysis is compatible with the observations of Komatsu~\cite{Komatsu2015} and Downs~\cite{Downs2021}, both being performed in a horizontal monolayer lattice vibrated off-plane.
In the former study~\cite{Komatsu2015}, the change in the nature of the liquid-solid transition, from a two-step continuous one to a discontinuous first-order-like one, was attributed to the particle inelasticity. In our case of a vertical monolayer, strongly inelastic collisions with the container's occur~\cite{Masmoudi2016} due to the inelastic collapse of the piles of grains~\cite{McNamara1992}.
In the latter study~\cite{Downs2021}, the bottom surface was decorated with periodically spaced dimples; the surface topography was also shown to alter the order of the phase transition and lead to a first-order one.


\section*{Conclusion}

In this work, we have studied the fluidization of a vertical monodisperse two-dimensional granular medium by increasing the energy injection and by quantifying what happens when modifying (and therefore disordering) the container’s geometry. We reported that altering the shape of the system favors the creation of disclinations and dislocations, breaking the hexagonal symmetry of the inner particulate lattice.
From image analysis, we obtained the instantaneous coordination number $Z$, compaction $C$, order parameter $\psi_6$ of each particle as a function of time. From this, we established a criterion to determine if a particle is in a solid-like or in a fluid-like phase, by analyzing the probability density function of the order parameter. We then performed a statistical analysis of the vibrated lattice, where we were able to tune the fluidized state by acting either on the base angle and/or the injected energy, independently. Along with this, we found that the velocity fluctuations tend to an isotropic-like Maxwell-Boltzmann distribution for configurations that favor disorder in the system. Finally, by analyzing the orientational order parameter susceptibility, we found that the system follows a first-order transition, triggered at $T=0$, between solid-like and fluid-like coexisting phases. A qualitative description then allowed us to bridge the thermal agitation and the solid-to-fluid fraction to the energy necessary to initiate a topological defect in the lattice.
In practice, our finding ends up with a way to control disorder and agitation by acting only on the geometry of the container, at a reduced injected energy cost.
In light of these observations, some points deserve to be addressed in future works, for instance by considering the benefits of a random container's shape instead of the single geometrical singularity of a V-shape base, if one wants to optimize further the thermal agitation at constant injected energy.
Such randomness includes considering the effect of rough walls, with typical surface roughness of the order of a fraction of beads' diameter, e.g. $R_q\sim 0.16\times d$, in agreement with our estimate of the activation energy of topological defects.
Another question is to probe the robustness and the limitations of our findings with regard to the polydispersity of the particles and to the relative size between the box and the particles, the effect of the geometry likely becoming less effective in both cases.
Also, the melting process here described is specific to 2D lattices but topological defects exist in all dimensions, suggesting a conceivable extrapolation of our findings. Another practical limitation of our system, in terms of acceleration, is the gravity below which very little happens. Here, a recent experiment conducted in a 3D sheared granular system~\cite{Xing2021} showed that topological defects, revealed by x-ray tomography, can be activated by the fluctuations of the container's volume, alike the more ordinary thermal agitation. This leads to the perspective of probing incommensurately small excitations, by analogy, in addition to determining up to what extent the shape of the container affects the mechanical response of granular media, in a general sense.



\section*{Methods}


\textit{Experimental Setup.\----}
The experimental setup is shown in Fig.~\ref{fig:setup_static}(a). It consists of a vertical Hele-Shaw cell filled with stainless steel beads (number $N=550$, diameter $d=2.0$~mm, density $\rho=8050$~kg/m$^3$, mass $m=33.7$~$\mu$g). The rectangular cell ($9 \times 18$ cm, gap $e=2.1$ mm) is made of aluminum and has a transparent acrylic front cover allowing the visualization of the grains in between. The gap is large enough to ensure that the particles freely move without being held between the front and back plates, but thin enough so that the out-of-plane tilt angle, $\theta_z\sim\sin^{-1}{[(e-d)/d]}\approx 3^{\circ}$, is sufficiently small to ensure that the layer of grains remains flat. At the bottom of the cell, we used a V-shape plastic template produced by a 3D printer (Zortrax model M200) using Z-ABS filament, for seven different angles $\theta = [0, 5, 8, 10, 20, 26, 30^{\circ}]$. A sample is prepared from a random rain-like deposit of particles in the cell. Due to the use of monodisperse spheres, the sample tends to organize, at rest, according to an equilateral triangular elementary cell to minimize the global potential energy~\cite{Andreotti2013}. In this sense, the range of V-shape angles, $\pi-2\theta$, fits within the limits of a hexagonal lattice. The features and repeatability of the initial static configurations are probed by reproducing ten times the rain-like deposit for each angle. The cell is then vibrated vertically by a shaker (Vibration Test Systems model VG 100-8 with power amplifier Techron Crown model 5515). The sinusoidal motion is given by $y(t)=A\sin(\omega t)$, where the frequency $f=\omega/2\pi=30$~Hz is kept constant in all our experiments. The dimensionless acceleration $\Gamma=A\omega^2/g$ is varied from $4$ to $10$, being $g$ the gravitational acceleration. An accelerometer (PCB Piezotronics model 352C03) is fixed at the top of the cell. The signal is then amplified by a signal conditioner (PCB Piezotronics model 480C02) and recorded by an oscilloscope (Tektronix model TDS 2014C) with a sample rate of $1$~kS/s. Finally, the cell is illuminated from the front by LED lights, and we track the position of the particles with a fast camera (Phantom, model M100) at $1600$~fps with a spatial resolution of $14$~pixels/mm during 4 seconds, i.e., over 120 driving periods.


\textit{Particle detection, velocity field and lattice topology.\----}
The reflection of LED light on the surface of the spherical particles produces a bright white spot in addition to smaller lateral spots due to the internal reflections on neighboring grains, see Fig.~\ref{fig:setup_static}(b). Spots with a minimal size are detected using IDL's particle tracking routines~\cite{Blair2008}, delivering a sub-pixel resolution of each spot's centroid. This procedure automatically discards the smaller secondary reflections and other spurious light sources. Knowing the positions of every particle, one can then obtain their trajectories as a function of time by minimizing the distance between pairs of positions in sequences of images. An example of the particle velocity field, obtained from the finite displacement of particles between two successive images, is shown in Fig.~\ref{fig:setup_static}(b). From the particle positions, the Delaunay triangulation (DT) of the lattice can also be calculated in every image. This representation gives information about the nearest neighbors (NN) of every particle. First, one can estimate the coordination number $Z(j)$, defined as the number of NN per each particle $j$. Then, one can quantify the local symmetry. A two-dimensional granular solid can indeed organize spontaneously into a hexagonal lattice~\cite{Buttinoni2017,Kansal2002} or into a square lattice~\cite{Castillo2012,Luu2013}. Within all of these, we checked that our vertical medium always favored the hexagonal arrangement for all the probed V-shape angles and driving accelerations (see for instance Fig.~\ref{fig:setup_static}(e) and the Supplementary Information). Therefore, we use the hexagonal symmetry as a reference, to describe a local order parameter~\cite{Chakrabarti1998} defined as $\psi_6(j) = \sum_{n=1}^{N_j} \exp(6i\theta_n^j)/N_j$, where $\theta_n^j$ corresponds to the angles with respect to the horizontal axis of the $n$-th NN of the $j$-th particle, as shown in Fig.\ref{fig:setup_static}(d). For example, in a perfect hexagonal close packing lattice, $|\psi_6^{hcp}|=1$, whereas $|\psi_6|\sim 0.5$ or below for an amorphous structure~\cite{Buttinoni2017}. Furthermore, the dual representation of the DT is the Voronoi tessellation (VT) represented in Fig.~\ref{fig:setup_static}(c). A single cell of the VT is a polygon whose edges are the perpendicular bisector of the DT segments. By construction, a Voronoi cell contains one particle only and delimits a region for one specific particle. This quantity allows us to define the local compaction $C(j)$, defined as the ratio between the particle's cross-sectional area to the VT cell area, $C(j)=\pi r^2/A_{VT}(j)$, where $r=d/2$ is the particle radius, and $A_{VT}(j)$ is the area of the polygonal VT cell~\cite{Rycroft2006} enclosing the $j$-th particle. In a two-dimensional hexagonal close packing lattice $A_{VT}^{hcp}=2\sqrt{3} r^2$ such that $C_{hcp}=\pi/2\sqrt{3}\simeq0.91$. We thus refer conveniently to a compaction relative to the hexagonal close packing, $C_r(j)=C(j)/C_{hcp}$ such that $0<C_r\leq1$. Note that in the maps shown in Fig.~\ref{fig:setup_static} and elsewhere in this study, the particles pertaining to the frontier of the sample are not considered in the ensemble averages to avoid biasing the estimations of the mean $C_r$ and $\psi_6$ with particles having arbitrary low $Z$.




\section*{Acknowledgments}

The authors thank Francisco Melo and Jean-Christophe G\'eminard for many fruitful discussions during the preparation of this work, Jean-Yves Choley for the support during R.Z.'s PhD thesis, and the CNRS's Laboratoire International Associ\'e `Mati\'ere, Structure et Dynamique' (LIA-MSD) in support of the Chile/France collaboration. R.Z. acknowledges ANID National Doctoral Program grant No. 21161404, the CY Cergy Paris Universit\'es Doctoral School 417 Sciences et Ing\'enierie, and the support of ISAE-Supm\'eca during his stay in France.\\

\noindent This version of the Article has been accepted for publication, after peer review but is not the Version of Record. The Version of Record is available online at \href{https://doi.org/10.1038/s41598-022-18965-4}{https://doi.org/10.1038/s41598-022-18965-4}.


\section*{Author contributions statement}

R.Z., G.V. and S.J. conceived the original idea, the methodology and the experiments, R.Z. conducted the experiments, G.V. and S.J. supervised the project, G.V. prepared the figures, S.J. analyzed and reviewed the dataset. All authors have discussed the results and contributed to both the manuscript and the Supplementary Information.


\section*{Additional information}
 
\textbf{Data availability statement}\\
The datasets generated and analysed during the current study are available from the corresponding author on reasonable request.\\
\textbf{Competing interests}\\
The authors declare no competing interests.



\begin{thebibliography}{10}
\urlstyle{rm}
\expandafter\ifx\csname url\endcsname\relax
  \def\url#1{\texttt{#1}}\fi
\expandafter\ifx\csname urlprefix\endcsname\relax\def\urlprefix{URL }\fi
\expandafter\ifx\csname doiprefix\endcsname\relax\def\doiprefix{DOI: }\fi
\providecommand{\bibinfo}[2]{#2}
\providecommand{\eprint}[2][]{\url{#2}}

\bibitem{Aranson2006}
\bibinfo{author}{Aranson, I.~S.} \& \bibinfo{author}{Tsimring, L.~S.}
\newblock \bibinfo{journal}{\bibinfo{title}{Patterns and collective behavior in
  granular media: Theoretical concepts}}.
\newblock {\emph{\JournalTitle{Rev. Mod. Phys.}}}
  \textbf{\bibinfo{volume}{78}}, \bibinfo{pages}{641--692},
  \doiprefix\url{10.1103/RevModPhys.78.641} (\bibinfo{year}{2006}).

\bibitem{jaeger1996}
\bibinfo{author}{Jaeger, H.~M.}, \bibinfo{author}{Nagel, S.~R.} \&
  \bibinfo{author}{Behringer, R.~P.}
\newblock \bibinfo{journal}{\bibinfo{title}{Granular solids, liquids, and
  gases}}.
\newblock {\emph{\JournalTitle{Rev. Mod. Phys.}}}
  \textbf{\bibinfo{volume}{68}}, \bibinfo{pages}{1259--1273},
  \doiprefix\url{10.1103/RevModPhys.68.1259} (\bibinfo{year}{1996}).

\bibitem{deGennes1999}
\bibinfo{author}{de~Gennes, P.~G.}
\newblock \bibinfo{journal}{\bibinfo{title}{Granular matter: a tentative
  view}}.
\newblock {\emph{\JournalTitle{Rev. Mod. Phys.}}}
  \textbf{\bibinfo{volume}{71}}, \bibinfo{pages}{S374--S382},
  \doiprefix\url{10.1103/RevModPhys.71.S374} (\bibinfo{year}{1999}).

\bibitem{Olafsen1998}
\bibinfo{author}{Olafsen, J.~S.} \& \bibinfo{author}{Urbach, J.~S.}
\newblock \bibinfo{journal}{\bibinfo{title}{Clustering, order, and collapse in
  a driven granular monolayer}}.
\newblock {\emph{\JournalTitle{Phys. Rev. Lett.}}}
  \textbf{\bibinfo{volume}{81}}, \bibinfo{pages}{4369--4372},
  \doiprefix\url{10.1103/physrevlett.81.4369} (\bibinfo{year}{1998}).

\bibitem{Quinn2000}
\bibinfo{author}{Quinn, P.~V.} \& \bibinfo{author}{Hong, D.~C.}
\newblock \bibinfo{journal}{\bibinfo{title}{Liquid-solid transition of hard
  spheres under gravity}}.
\newblock {\emph{\JournalTitle{Phys. Rev. E}}} \textbf{\bibinfo{volume}{62}},
  \bibinfo{pages}{8295--8298}, \doiprefix\url{10.1103/physreve.62.8295}
  (\bibinfo{year}{2000}).

\bibitem{Falcon2001}
\bibinfo{author}{Falcon, {\'{E}}.}, \bibinfo{author}{Fauve, S.} \&
  \bibinfo{author}{Laroche, C.}
\newblock \bibinfo{title}{Experimental study of a granular gas fluidized by
  vibrations}.
\newblock In \emph{\bibinfo{booktitle}{Granular Gases}},
  \bibinfo{pages}{244--253}, \doiprefix\url{10.1007/3-540-44506-4_14}
  (\bibinfo{publisher}{Springer Berlin Heidelberg}, \bibinfo{year}{2001}).

\bibitem{Olafsen2005}
\bibinfo{author}{Olafsen, J.~S.} \& \bibinfo{author}{Urbach, J.~S.}
\newblock \bibinfo{journal}{\bibinfo{title}{Two-dimensional melting far from
  equilibrium in a granular monolayer}}.
\newblock {\emph{\JournalTitle{Phys. Rev. Lett.}}}
  \textbf{\bibinfo{volume}{95}}, \bibinfo{pages}{098002},
  \doiprefix\url{10.1103/physrevlett.95.098002} (\bibinfo{year}{2005}).

\bibitem{Daniels2006}
\bibinfo{author}{Daniels, K.~E.} \& \bibinfo{author}{Behringer, R.~P.}
\newblock \bibinfo{journal}{\bibinfo{title}{Characterization of a
  freezing/melting transition in a vibrated and sheared granular medium}}.
\newblock {\emph{\JournalTitle{J. Stat. Mech: Theory Exp.}}}
  \textbf{\bibinfo{volume}{2006}}, \bibinfo{pages}{P07018--P07018},
  \doiprefix\url{10.1088/1742-5468/2006/07/p07018} (\bibinfo{year}{2006}).

\bibitem{Reis2006}
\bibinfo{author}{Reis, P.~M.}, \bibinfo{author}{Ingale, R.~A.} \&
  \bibinfo{author}{Shattuck, M.~D.}
\newblock \bibinfo{journal}{\bibinfo{title}{Crystallization of a
  quasi-two-dimensional granular fluid}}.
\newblock {\emph{\JournalTitle{Phys. Rev. Lett.}}}
  \textbf{\bibinfo{volume}{96}}, \bibinfo{pages}{258001},
  \doiprefix\url{10.1103/PhysRevLett.96.258001} (\bibinfo{year}{2006}).

\bibitem{Roeller2011}
\bibinfo{author}{Roeller, K.}, \bibinfo{author}{Clewett, J. P.~D.},
  \bibinfo{author}{Bowley, R.~M.}, \bibinfo{author}{Herminghaus, S.} \&
  \bibinfo{author}{Swift, M.~R.}
\newblock \bibinfo{journal}{\bibinfo{title}{Liquid-gas phase separation in
  confined vibrated dry granular matter}}.
\newblock {\emph{\JournalTitle{Phys. Rev. Lett.}}}
  \textbf{\bibinfo{volume}{107}}, \bibinfo{pages}{048002},
  \doiprefix\url{10.1103/physrevlett.107.048002} (\bibinfo{year}{2011}).

\bibitem{Heckel2015}
\bibinfo{author}{Heckel, M.}, \bibinfo{author}{Sack, A.},
  \bibinfo{author}{Kollmer, J.~E.} \& \bibinfo{author}{P\"oschel, T.}
\newblock \bibinfo{journal}{\bibinfo{title}{Fluidization of a horizontally
  driven granular monolayer}}.
\newblock {\emph{\JournalTitle{Physical Review E}}}
  \textbf{\bibinfo{volume}{91}}, \bibinfo{pages}{062213},
  \doiprefix\url{10.1103/physreve.91.062213} (\bibinfo{year}{2015}).

\bibitem{Clewett2016}
\bibinfo{author}{Clewett, J. P.~D.} \emph{et~al.}
\newblock \bibinfo{journal}{\bibinfo{title}{The minimization of mechanical work
  in vibrated granular matter}}.
\newblock {\emph{\JournalTitle{Scientific Reports}}}
  \textbf{\bibinfo{volume}{6}}, \bibinfo{pages}{28726},
  \doiprefix\url{10.1038/srep28726} (\bibinfo{year}{2016}).

\bibitem{Clewett2012}
\bibinfo{author}{Clewett, J. P.~D.}, \bibinfo{author}{Roeller, K.},
  \bibinfo{author}{Bowley, R.~M.}, \bibinfo{author}{Herminghaus, S.} \&
  \bibinfo{author}{Swift, M.~R.}
\newblock \bibinfo{journal}{\bibinfo{title}{Emergent surface tension in
  vibrated, noncohesive granular media}}.
\newblock {\emph{\JournalTitle{Phys. Rev. Lett.}}}
  \textbf{\bibinfo{volume}{109}}, \bibinfo{pages}{228002},
  \doiprefix\url{10.1103/physrevlett.109.228002} (\bibinfo{year}{2012}).

\bibitem{Noirhomme2021}
\bibinfo{author}{Noirhomme, M.} \emph{et~al.}
\newblock \bibinfo{journal}{\bibinfo{title}{Particle dynamics at the onset of
  the granular gas-liquid transition}}.
\newblock {\emph{\JournalTitle{Phys. Rev. Lett.}}}
  \textbf{\bibinfo{volume}{126}}, \bibinfo{pages}{128002},
  \doiprefix\url{10.1103/physrevlett.126.128002} (\bibinfo{year}{2021}).

\bibitem{Melo1995}
\bibinfo{author}{Melo, F.}, \bibinfo{author}{Umbanhowar, P.~B.} \&
  \bibinfo{author}{Swinney, H.~L.}
\newblock \bibinfo{journal}{\bibinfo{title}{Hexagons, kinks, and disorder in
  oscillated granular layers}}.
\newblock {\emph{\JournalTitle{Phys. Rev. Lett.}}}
  \textbf{\bibinfo{volume}{75}}, \bibinfo{pages}{3838--3841},
  \doiprefix\url{10.1103/physrevlett.75.3838} (\bibinfo{year}{1995}).

\bibitem{Umbanhowar1996}
\bibinfo{author}{Umbanhowar, P.~B.}, \bibinfo{author}{Melo, F.} \&
  \bibinfo{author}{Swinney, H.~L.}
\newblock \bibinfo{journal}{\bibinfo{title}{Localized excitations in a
  vertically vibrated granular layer}}.
\newblock {\emph{\JournalTitle{Nature}}} \textbf{\bibinfo{volume}{382}},
  \bibinfo{pages}{793--796}, \doiprefix\url{10.1038/382793a0}
  (\bibinfo{year}{1996}).

\bibitem{Douady1989}
\bibinfo{author}{Douady, S.}, \bibinfo{author}{Fauve, S.} \&
  \bibinfo{author}{Laroche, C.}
\newblock \bibinfo{journal}{\bibinfo{title}{Subharmonic instabilities and
  defects in a granular layer under vertical vibrations}}.
\newblock {\emph{\JournalTitle{Europhysics Letters}}}
  \textbf{\bibinfo{volume}{8}}, \bibinfo{pages}{621--627},
  \doiprefix\url{10.1209/0295-5075/8/7/007} (\bibinfo{year}{1989}).

\bibitem{Theocharis2009}
\bibinfo{author}{Theocharis, G.} \emph{et~al.}
\newblock \bibinfo{journal}{\bibinfo{title}{Localized breathing modes in
  granular crystals with defects}}.
\newblock {\emph{\JournalTitle{Physical Review E}}}
  \textbf{\bibinfo{volume}{80}}, \bibinfo{pages}{066601},
  \doiprefix\url{10.1103/physreve.80.066601} (\bibinfo{year}{2009}).

\bibitem{Job2009}
\bibinfo{author}{Job, S.}, \bibinfo{author}{Santibanez, F.},
  \bibinfo{author}{Tapia, F.} \& \bibinfo{author}{Melo, F.}
\newblock \bibinfo{journal}{\bibinfo{title}{Wave localization in strongly
  nonlinear hertzian chains with mass defect}}.
\newblock {\emph{\JournalTitle{Physical Review E}}}
  \textbf{\bibinfo{volume}{80}}, \bibinfo{pages}{025602},
  \doiprefix\url{10.1103/physreve.80.025602} (\bibinfo{year}{2009}).

\bibitem{Boechler2010}
\bibinfo{author}{Boechler, N.} \emph{et~al.}
\newblock \bibinfo{journal}{\bibinfo{title}{Discrete breathers in
  one-dimensional diatomic granular crystals}}.
\newblock {\emph{\JournalTitle{Phys. Rev. Lett.}}}
  \textbf{\bibinfo{volume}{104}}, \bibinfo{pages}{244302},
  \doiprefix\url{10.1103/PhysRevLett.104.244302} (\bibinfo{year}{2010}).

\bibitem{Ponson2010}
\bibinfo{author}{Ponson, L.} \emph{et~al.}
\newblock \bibinfo{journal}{\bibinfo{title}{Nonlinear waves in disordered
  diatomic granular chains}}.
\newblock {\emph{\JournalTitle{Phys. Rev. E}}} \textbf{\bibinfo{volume}{82}},
  \bibinfo{pages}{021301}, \doiprefix\url{10.1103/PhysRevE.82.021301}
  (\bibinfo{year}{2010}).

\bibitem{Briand2016}
\bibinfo{author}{Briand, G.} \& \bibinfo{author}{Dauchot, O.}
\newblock \bibinfo{journal}{\bibinfo{title}{Crystallization of self-propelled
  hard discs}}.
\newblock {\emph{\JournalTitle{Phys. Rev. Lett.}}}
  \textbf{\bibinfo{volume}{117}}, \bibinfo{pages}{098004},
  \doiprefix\url{10.1103/physrevlett.117.098004} (\bibinfo{year}{2016}).

\bibitem{Klamser2018}
\bibinfo{author}{Klamser, J.~U.}, \bibinfo{author}{Kapfer, S.~C.} \&
  \bibinfo{author}{Krauth, W.}
\newblock \bibinfo{journal}{\bibinfo{title}{Thermodynamic phases in
  two-dimensional active matter}}.
\newblock {\emph{\JournalTitle{Nat. Commun.}}} \textbf{\bibinfo{volume}{9}},
  \bibinfo{pages}{5045}, \doiprefix\url{10.1038/s41467-018-07491-5}
  (\bibinfo{year}{2018}).

\bibitem{Rietz2018}
\bibinfo{author}{Rietz, F.}, \bibinfo{author}{Radin, C.},
  \bibinfo{author}{Swinney, H.~L.} \& \bibinfo{author}{Schröter, M.}
\newblock \bibinfo{journal}{\bibinfo{title}{Nucleation in sheared granular
  matter}}.
\newblock {\emph{\JournalTitle{Phys. Rev. Lett.}}}
  \textbf{\bibinfo{volume}{120}}, \bibinfo{pages}{055701},
  \doiprefix\url{10.1103/physrevlett.120.055701} (\bibinfo{year}{2018}).

\bibitem{Strandburg1988}
\bibinfo{author}{Strandburg, K.~J.}
\newblock \bibinfo{journal}{\bibinfo{title}{Two-dimensional melting}}.
\newblock {\emph{\JournalTitle{Rev. Mod. Phys.}}}
  \textbf{\bibinfo{volume}{60}}, \bibinfo{pages}{161--207},
  \doiprefix\url{10.1103/RevModPhys.60.161} (\bibinfo{year}{1988}).

\bibitem{Kosterlitz2017}
\bibinfo{author}{Kosterlitz, J.~M.}
\newblock \bibinfo{journal}{\bibinfo{title}{Nobel lecture: Topological defects
  and phase transitions}}.
\newblock {\emph{\JournalTitle{Rev. Mod. Phys.}}}
  \textbf{\bibinfo{volume}{89}}, \bibinfo{pages}{040501},
  \doiprefix\url{10.1103/revmodphys.89.040501} (\bibinfo{year}{2017}).

\bibitem{Kosterlitz1972}
\bibinfo{author}{Kosterlitz, J.~M.} \& \bibinfo{author}{Thouless, D.~J.}
\newblock \bibinfo{journal}{\bibinfo{title}{Long range order and metastability
  in two dimensional solids and superfluids. (application of dislocation
  theory)}}.
\newblock {\emph{\JournalTitle{J. Phys. C: Solid State Phys.}}}
  \textbf{\bibinfo{volume}{5}}, \bibinfo{pages}{L124--L126},
  \doiprefix\url{10.1088/0022-3719/5/11/002} (\bibinfo{year}{1972}).

\bibitem{Kosterlitz1973}
\bibinfo{author}{Kosterlitz, J.~M.} \& \bibinfo{author}{Thouless, D.~J.}
\newblock \bibinfo{journal}{\bibinfo{title}{Ordering, metastability and phase
  transitions in two-dimensional systems}}.
\newblock {\emph{\JournalTitle{J. Phys. C: Solid State Phys.}}}
  \textbf{\bibinfo{volume}{6}}, \bibinfo{pages}{1181--1203},
  \doiprefix\url{10.1088/0022-3719/6/7/010} (\bibinfo{year}{1973}).

\bibitem{Berthier2011}
\bibinfo{author}{Berthier, L.} \& \bibinfo{author}{Biroli, G.}
\newblock \bibinfo{journal}{\bibinfo{title}{Theoretical perspective on the
  glass transition and amorphous materials}}.
\newblock {\emph{\JournalTitle{Rev. Mod. Phys.}}}
  \textbf{\bibinfo{volume}{83}}, \bibinfo{pages}{587--645},
  \doiprefix\url{10.1103/revmodphys.83.587} (\bibinfo{year}{2011}).

\bibitem{Chen2021}
\bibinfo{author}{Chen, Y.} \emph{et~al.}
\newblock \bibinfo{journal}{\bibinfo{title}{2d colloidal crystals with
  anisotropic impurities}}.
\newblock {\emph{\JournalTitle{Phys. Rev. Lett.}}}
  \textbf{\bibinfo{volume}{127}}, \bibinfo{pages}{018004},
  \doiprefix\url{10.1103/physrevlett.127.018004} (\bibinfo{year}{2021}).

\bibitem{Fortini2015}
\bibinfo{author}{Fortini, A.} \& \bibinfo{author}{Huang, K.}
\newblock \bibinfo{journal}{\bibinfo{title}{Role of defects in the onset of
  wall-induced granular convection}}.
\newblock {\emph{\JournalTitle{Phys. Rev. E}}} \textbf{\bibinfo{volume}{91}},
  \bibinfo{pages}{032206}, \doiprefix\url{10.1103/physreve.91.032206}
  (\bibinfo{year}{2015}).

\bibitem{Sun2016}
\bibinfo{author}{Sun, X.}, \bibinfo{author}{Li, Y.}, \bibinfo{author}{Ma, Y.}
  \& \bibinfo{author}{Zhang, Z.}
\newblock \bibinfo{journal}{\bibinfo{title}{Direct observation of melting in a
  two-dimensional driven granular system}}.
\newblock {\emph{\JournalTitle{Scientific Reports}}}
  \textbf{\bibinfo{volume}{6}}, \bibinfo{pages}{24056},
  \doiprefix\url{10.1038/srep24056} (\bibinfo{year}{2016}).

\bibitem{Ramming2017}
\bibinfo{author}{Ramming, P.} \& \bibinfo{author}{Huang, K.}
\newblock \bibinfo{journal}{\bibinfo{title}{Clustering and melting in a wet
  granular monolayer}}.
\newblock {\emph{\JournalTitle{EPJ Web Conf - Powders and Grains 2017}}}
  \textbf{\bibinfo{volume}{140}}, \bibinfo{pages}{08003},
  \doiprefix\url{10.1051/epjconf/201714008003} (\bibinfo{year}{2017}).

\bibitem{Thorneywork2017}
\bibinfo{author}{Thorneywork, A.~L.}, \bibinfo{author}{Abbott, J.~L.},
  \bibinfo{author}{Aarts, D.~G.} \& \bibinfo{author}{Dullens, R.~P.}
\newblock \bibinfo{journal}{\bibinfo{title}{Two-dimensional melting of
  colloidal hard spheres}}.
\newblock {\emph{\JournalTitle{Phys. Rev. Lett.}}}
  \textbf{\bibinfo{volume}{118}}, \bibinfo{pages}{158001},
  \doiprefix\url{10.1103/physrevlett.118.158001} (\bibinfo{year}{2017}).

\bibitem{Cao2018}
\bibinfo{author}{Cao, Y.} \emph{et~al.}
\newblock \bibinfo{journal}{\bibinfo{title}{Structural and topological nature
  of plasticity in sheared granular materials}}.
\newblock {\emph{\JournalTitle{Nat. Commun.}}} \textbf{\bibinfo{volume}{9}},
  \bibinfo{pages}{2911}, \doiprefix\url{10.1038/s41467-018-05329-8}
  (\bibinfo{year}{2018}).

\bibitem{Arjun2020}
\bibinfo{author}{H, A.} \& \bibinfo{author}{Chaudhuri, P.}
\newblock \bibinfo{journal}{\bibinfo{title}{Dense hard disk ordering: influence
  of bidispersity and quenched disorder}}.
\newblock {\emph{\JournalTitle{J. Phys.: Condens. Matter}}}
  \textbf{\bibinfo{volume}{32}}, \bibinfo{pages}{414001},
  \doiprefix\url{10.1088/1361-648x/ab9b52} (\bibinfo{year}{2020}).

\bibitem{Binder2002}
\bibinfo{author}{Binder, K.}, \bibinfo{author}{Sengupta, S.} \&
  \bibinfo{author}{Nielaba, P.}
\newblock \bibinfo{journal}{\bibinfo{title}{The liquid-solid transition of hard
  discs: first-order transition or kosterlitz-thouless-halperin-nelson-young
  scenario?}}
\newblock {\emph{\JournalTitle{J. Phys.: Condens. Matter}}}
  \textbf{\bibinfo{volume}{14}}, \bibinfo{pages}{2323--2333},
  \doiprefix\url{10.1088/0953-8984/14/9/321} (\bibinfo{year}{2002}).

\bibitem{Tuckman2020}
\bibinfo{author}{Tuckman, P.~J.} \emph{et~al.}
\newblock \bibinfo{journal}{\bibinfo{title}{Contact network changes in ordered
  and disordered disk packings}}.
\newblock {\emph{\JournalTitle{Soft Matter}}} \textbf{\bibinfo{volume}{16}},
  \bibinfo{pages}{9443--9455}, \doiprefix\url{10.1039/d0sm01137a}
  (\bibinfo{year}{2020}).

\bibitem{Prevost2004}
\bibinfo{author}{Prevost, A.}, \bibinfo{author}{Melby, P.},
  \bibinfo{author}{Egolf, D.~A.} \& \bibinfo{author}{Urbach, J.~S.}
\newblock \bibinfo{journal}{\bibinfo{title}{Nonequilibrium two-phase
  coexistence in a confined granular layer}}.
\newblock {\emph{\JournalTitle{Phys. Rev. E}}} \textbf{\bibinfo{volume}{70}},
  \bibinfo{pages}{050301}, \doiprefix\url{10.1103/physreve.70.050301}
  (\bibinfo{year}{2004}).

\bibitem{Goetzendorfer2005}
\bibinfo{author}{G{\"o}tzendorfer, A.}, \bibinfo{author}{Kreft, J.},
  \bibinfo{author}{Kruelle, C.~A.} \& \bibinfo{author}{Rehberg, I.}
\newblock \bibinfo{journal}{\bibinfo{title}{Sublimation of a vibrated granular
  monolayer: Coexistence of gas and solid}}.
\newblock {\emph{\JournalTitle{Phys. Rev. Lett.}}}
  \textbf{\bibinfo{volume}{95}}, \bibinfo{pages}{135704},
  \doiprefix\url{10.1103/physrevlett.95.135704} (\bibinfo{year}{2005}).

\bibitem{Melby2005}
\bibinfo{author}{Melby, P.} \emph{et~al.}
\newblock \bibinfo{journal}{\bibinfo{title}{The dynamics of thin vibrated
  granular layers}}.
\newblock {\emph{\JournalTitle{J. Phys.: Condens. Matter}}}
  \textbf{\bibinfo{volume}{17}}, \bibinfo{pages}{S2689--S2704},
  \doiprefix\url{10.1088/0953-8984/17/24/020} (\bibinfo{year}{2005}).

\bibitem{Pastor2007}
\bibinfo{author}{Pastor, J.}, \bibinfo{author}{Maza, D.},
  \bibinfo{author}{Zuriguel, I.}, \bibinfo{author}{Garcimart{\'{\i}}n, A.} \&
  \bibinfo{author}{Boudet, J.-F.}
\newblock \bibinfo{journal}{\bibinfo{title}{Time resolved particle dynamics in
  granular convection}}.
\newblock {\emph{\JournalTitle{Physica D}}} \textbf{\bibinfo{volume}{232}},
  \bibinfo{pages}{128--135}, \doiprefix\url{10.1016/j.physd.2007.06.005}
  (\bibinfo{year}{2007}).

\bibitem{Bernard2011}
\bibinfo{author}{Bernard, E.~P.} \& \bibinfo{author}{Krauth, W.}
\newblock \bibinfo{journal}{\bibinfo{title}{Two-step melting in two dimensions:
  First-order liquid-hexatic transition}}.
\newblock {\emph{\JournalTitle{Physical Review Letters}}}
  \textbf{\bibinfo{volume}{107}}, \bibinfo{pages}{155704},
  \doiprefix\url{10.1103/physrevlett.107.155704} (\bibinfo{year}{2011}).

\bibitem{Castillo2012}
\bibinfo{author}{Castillo, G.}, \bibinfo{author}{Mujica, N.} \&
  \bibinfo{author}{Soto, R.}
\newblock \bibinfo{journal}{\bibinfo{title}{Fluctuations and criticality of a
  granular solid-liquid-like phase transition}}.
\newblock {\emph{\JournalTitle{Phys. Rev. Lett.}}}
  \textbf{\bibinfo{volume}{109}}, \bibinfo{pages}{095701},
  \doiprefix\url{10.1103/physrevlett.109.095701} (\bibinfo{year}{2012}).

\bibitem{Qi2014}
\bibinfo{author}{Qi, W.}, \bibinfo{author}{Gantapara, A.~P.} \&
  \bibinfo{author}{Dijkstra, M.}
\newblock \bibinfo{journal}{\bibinfo{title}{Two-stage melting induced by
  dislocations and grain boundaries in monolayers of hard spheres}}.
\newblock {\emph{\JournalTitle{Soft Matter}}} \textbf{\bibinfo{volume}{10}},
  \bibinfo{pages}{5449}, \doiprefix\url{10.1039/c4sm00125g}
  (\bibinfo{year}{2014}).

\bibitem{Kapfer2015}
\bibinfo{author}{Kapfer, S.~C.} \& \bibinfo{author}{Krauth, W.}
\newblock \bibinfo{journal}{\bibinfo{title}{Two-dimensional melting: From
  liquid-hexatic coexistence to continuous transitions}}.
\newblock {\emph{\JournalTitle{Physical Review Letters}}}
  \textbf{\bibinfo{volume}{114}}, \bibinfo{pages}{035702},
  \doiprefix\url{10.1103/physrevlett.114.035702} (\bibinfo{year}{2015}).

\bibitem{Mujica2016}
\bibinfo{author}{Mujica, N.} \& \bibinfo{author}{Soto, R.}
\newblock \bibinfo{title}{Dynamics of noncohesive confined granular media}.
\newblock In \bibinfo{editor}{Klapp, J.}, \bibinfo{editor}{Sigalotti, L.
  D.~G.}, \bibinfo{editor}{Medina, A.}, \bibinfo{editor}{L{\'o}pez, A.} \&
  \bibinfo{editor}{Ruiz-Chavarr{\'i}a, G.} (eds.)
  \emph{\bibinfo{booktitle}{Recent Advances in Fluid Dynamics with
  Environmental Applications}}, \bibinfo{pages}{445--463},
  \doiprefix\url{10.1007/978-3-319-27965-7_32} (\bibinfo{publisher}{Springer
  International Publishing}, \bibinfo{year}{2016}).

\bibitem{Clerc2008}
\bibinfo{author}{Clerc, M.~G.} \emph{et~al.}
\newblock \bibinfo{journal}{\bibinfo{title}{Liquid{\textendash}solid-like
  transition in quasi-one-dimensional driven granular media}}.
\newblock {\emph{\JournalTitle{Nat. Phys.}}} \textbf{\bibinfo{volume}{4}},
  \bibinfo{pages}{249--254}, \doiprefix\url{10.1038/nphys884}
  (\bibinfo{year}{2008}).

\bibitem{Komatsu2015}
\bibinfo{author}{Komatsu, Y.} \& \bibinfo{author}{Tanaka, H.}
\newblock \bibinfo{journal}{\bibinfo{title}{Roles of energy dissipation in a
  liquid-solid transition of out-of-equilibrium systems}}.
\newblock {\emph{\JournalTitle{Phys. Rev. X}}} \textbf{\bibinfo{volume}{5}},
  \bibinfo{pages}{031025}, \doiprefix\url{10.1103/physrevx.5.031025}
  (\bibinfo{year}{2015}).

\bibitem{Downs2021}
\bibinfo{author}{Downs, J.~G.}, \bibinfo{author}{Smith, N.~D.},
  \bibinfo{author}{Mandadapu, K.~K.}, \bibinfo{author}{Garrahan, J.~P.} \&
  \bibinfo{author}{Smith, M.~I.}
\newblock \bibinfo{journal}{\bibinfo{title}{Topographic control of order in
  quasi-2d granular phase transitions}}.
\newblock {\emph{\JournalTitle{Physical Review Letters}}}
  \textbf{\bibinfo{volume}{127}}, \bibinfo{pages}{268002},
  \doiprefix\url{10.1103/physrevlett.127.268002} (\bibinfo{year}{2021}).

\bibitem{Ottino2000}
\bibinfo{author}{Ottino, J.~M.} \& \bibinfo{author}{Khakhar, D.~V.}
\newblock \bibinfo{journal}{\bibinfo{title}{Mixing and segregation of granular
  materials}}.
\newblock {\emph{\JournalTitle{Annual Review of Fluid Mechanics}}}
  \textbf{\bibinfo{volume}{32}}, \bibinfo{pages}{55--91},
  \doiprefix\url{10.1146/annurev.fluid.32.1.55} (\bibinfo{year}{2000}).

\bibitem{Barker2021}
\bibinfo{author}{Barker, T.}, \bibinfo{author}{Rauter, M.},
  \bibinfo{author}{Maguire, E. S.~F.}, \bibinfo{author}{Johnson, C.~G.} \&
  \bibinfo{author}{Gray, J. M. N.~T.}
\newblock \bibinfo{journal}{\bibinfo{title}{Coupling rheology and segregation
  in granular flows}}.
\newblock {\emph{\JournalTitle{Journal of Fluid Mechanics}}}
  \textbf{\bibinfo{volume}{909}}, \bibinfo{pages}{A22},
  \doiprefix\url{10.1017/jfm.2020.973} (\bibinfo{year}{2021}).

\bibitem{Saluena1999}
\bibinfo{author}{Salue\~na, C.}, \bibinfo{author}{P\"oschel, T.} \&
  \bibinfo{author}{Esipov, S.~E.}
\newblock \bibinfo{journal}{\bibinfo{title}{Dissipative properties of vibrated
  granular materials}}.
\newblock {\emph{\JournalTitle{Phys. Rev. E}}} \textbf{\bibinfo{volume}{59}},
  \bibinfo{pages}{4422--4425}, \doiprefix\url{10.1103/PhysRevE.59.4422}
  (\bibinfo{year}{1999}).

\bibitem{Sanchez2012}
\bibinfo{author}{S\'anchez, M.}, \bibinfo{author}{Rosenthal, G.} \&
  \bibinfo{author}{Pugnaloni, L.~A.}
\newblock \bibinfo{journal}{\bibinfo{title}{Universal response of optimal
  granular damping devices}}.
\newblock {\emph{\JournalTitle{J. Sound Vib.}}} \textbf{\bibinfo{volume}{331}},
  \bibinfo{pages}{4389--4394}, \doiprefix\url{10.1016/j.jsv.2012.05.001}
  (\bibinfo{year}{2012}).

\bibitem{Sack2013}
\bibinfo{author}{Sack, A.}, \bibinfo{author}{Heckel, M.},
  \bibinfo{author}{Kollmer, J.~E.}, \bibinfo{author}{Zimber, F.} \&
  \bibinfo{author}{P\"oschel, T.}
\newblock \bibinfo{journal}{\bibinfo{title}{Energy dissipation in driven
  granular matter in the absence of gravity}}.
\newblock {\emph{\JournalTitle{Phys. Rev. Lett.}}}
  \textbf{\bibinfo{volume}{111}}, \bibinfo{pages}{018001},
  \doiprefix\url{10.1103/PhysRevLett.111.018001} (\bibinfo{year}{2013}).

\bibitem{Masmoudi2016}
\bibinfo{author}{Masmoudi, M.}, \bibinfo{author}{Job, S.},
  \bibinfo{author}{Abbes, M.~S.}, \bibinfo{author}{Tawfiq, I.} \&
  \bibinfo{author}{Haddar, M.}
\newblock \bibinfo{journal}{\bibinfo{title}{Experimental and numerical
  investigations of dissipation mechanisms in particle dampers}}.
\newblock {\emph{\JournalTitle{Granular Matter}}}
  \textbf{\bibinfo{volume}{18}}, \bibinfo{pages}{71},
  \doiprefix\url{10.1007/s10035-016-0667-4} (\bibinfo{year}{2016}).

\bibitem{Ferreyra2021}
\bibinfo{author}{Ferreyra, M.~V.}, \bibinfo{author}{Baldini, M.},
  \bibinfo{author}{Pugnaloni, L.~A.} \& \bibinfo{author}{Job, S.}
\newblock \bibinfo{journal}{\bibinfo{title}{Effect of lateral confinement on
  the apparent mass of granular dampers}}.
\newblock {\emph{\JournalTitle{Granular Matter}}}
  \textbf{\bibinfo{volume}{23}}, \bibinfo{pages}{45},
  \doiprefix\url{10.1007/s10035-021-01090-w} (\bibinfo{year}{2021}).

\bibitem{Hattar2011}
\bibinfo{author}{Hattar, K.}
\newblock \bibinfo{title}{Deformation structures including twins in nanograined
  pure metals}.
\newblock In \emph{\bibinfo{booktitle}{Nanostructured Metals and Alloys}},
  \bibinfo{pages}{213--242}, \doiprefix\url{10.1533/9780857091123.2.213}
  (\bibinfo{publisher}{Elsevier}, \bibinfo{year}{2011}).

\bibitem{Shen2019}
\bibinfo{author}{Shen, H.}, \bibinfo{author}{Tong, H.}, \bibinfo{author}{Tan,
  P.} \& \bibinfo{author}{Xu, L.}
\newblock \bibinfo{journal}{\bibinfo{title}{A universal state and its
  relaxation mechanisms of long-range interacting polygons}}.
\newblock {\emph{\JournalTitle{Nature Communications}}}
  \textbf{\bibinfo{volume}{10}}, \bibinfo{pages}{1737},
  \doiprefix\url{10.1038/s41467-019-09795-6} (\bibinfo{year}{2019}).

\bibitem{Luu2013}
\bibinfo{author}{Luu, L.-H.}, \bibinfo{author}{Castillo, G.},
  \bibinfo{author}{Mujica, N.} \& \bibinfo{author}{Soto, R.}
\newblock \bibinfo{journal}{\bibinfo{title}{Capillarylike fluctuations of a
  solid-liquid interface in a noncohesive granular system}}.
\newblock {\emph{\JournalTitle{Phys. Rev. E}}} \textbf{\bibinfo{volume}{87}},
  \bibinfo{pages}{040202}, \doiprefix\url{10.1103/physreve.87.040202}
  (\bibinfo{year}{2013}).

\bibitem{Han2008}
\bibinfo{author}{Han, Y.}, \bibinfo{author}{Ha, N.~Y.},
  \bibinfo{author}{Alsayed, A.~M.} \& \bibinfo{author}{Yodh, A.~G.}
\newblock \bibinfo{journal}{\bibinfo{title}{Melting of two-dimensional
  tunable-diameter colloidal crystals}}.
\newblock {\emph{\JournalTitle{Physical Review E}}}
  \textbf{\bibinfo{volume}{77}}, \bibinfo{pages}{041406},
  \doiprefix\url{10.1103/physreve.77.041406} (\bibinfo{year}{2008}).

\bibitem{Buttinoni2017}
\bibinfo{author}{Buttinoni, I.} \emph{et~al.}
\newblock \bibinfo{journal}{\bibinfo{title}{Direct observation of impact
  propagation and absorption in dense colloidal monolayers}}.
\newblock {\emph{\JournalTitle{Proc. Natl. Acad. Sci. USA}}}
  \textbf{\bibinfo{volume}{114}}, \bibinfo{pages}{12150--12155},
  \doiprefix\url{10.1073/pnas.1712266114} (\bibinfo{year}{2017}).

\bibitem{Kansal2002}
\bibinfo{author}{Kansal, A.~R.}, \bibinfo{author}{Torquato, S.} \&
  \bibinfo{author}{Stillinger, F.~H.}
\newblock \bibinfo{journal}{\bibinfo{title}{Diversity of order and densities in
  jammed hard-particle packings}}.
\newblock {\emph{\JournalTitle{Phys. Rev. E}}} \textbf{\bibinfo{volume}{66}},
  \bibinfo{pages}{041109}, \doiprefix\url{10.1103/physreve.66.041109}
  (\bibinfo{year}{2002}).

\bibitem{Chakrabarti1998}
\bibinfo{author}{Chakrabarti, J.} \& \bibinfo{author}{L{\"o}wen, H.}
\newblock \bibinfo{journal}{\bibinfo{title}{Effect of confinement on
  charge-stabilized colloidal suspensions between two charged plates}}.
\newblock {\emph{\JournalTitle{Phys. Rev. E}}} \textbf{\bibinfo{volume}{58}},
  \bibinfo{pages}{3400--3404}, \doiprefix\url{10.1103/physreve.58.3400}
  (\bibinfo{year}{1998}).

\bibitem{Olafsen1999}
\bibinfo{author}{Olafsen, J.~S.} \& \bibinfo{author}{Urbach, J.~S.}
\newblock \bibinfo{journal}{\bibinfo{title}{Velocity distributions and density
  fluctuations in a granular gas}}.
\newblock {\emph{\JournalTitle{Phys. Rev. E}}} \textbf{\bibinfo{volume}{60}},
  \bibinfo{pages}{R2468--R2471}, \doiprefix\url{10.1103/physreve.60.r2468}
  (\bibinfo{year}{1999}).

\bibitem{Losert1999}
\bibinfo{author}{Losert, W.}, \bibinfo{author}{Cooper, D. G.~W.},
  \bibinfo{author}{Delour, J.}, \bibinfo{author}{Kudrolli, A.} \&
  \bibinfo{author}{Gollub, J.~P.}
\newblock \bibinfo{journal}{\bibinfo{title}{Velocity statistics in excited
  granular media}}.
\newblock {\emph{\JournalTitle{Chaos}}} \textbf{\bibinfo{volume}{9}},
  \bibinfo{pages}{682--690}, \doiprefix\url{10.1063/1.166442}
  (\bibinfo{year}{1999}).

\bibitem{Losert2000}
\bibinfo{author}{Losert, W.}, \bibinfo{author}{Bocquet, L.},
  \bibinfo{author}{Lubensky, T.~C.} \& \bibinfo{author}{Gollub, J.~P.}
\newblock \bibinfo{journal}{\bibinfo{title}{Particle dynamics in sheared
  granular matter}}.
\newblock {\emph{\JournalTitle{Phys. Rev. Lett.}}}
  \textbf{\bibinfo{volume}{85}}, \bibinfo{pages}{1428--1431},
  \doiprefix\url{10.1103/physrevlett.85.1428} (\bibinfo{year}{2000}).

\bibitem{Baxter2006}
\bibinfo{author}{Baxter, G.~W.} \& \bibinfo{author}{Olafsen, J.~S.}
\newblock \bibinfo{journal}{\bibinfo{title}{The temperature of a vibrated
  granular gas}}.
\newblock {\emph{\JournalTitle{Granular Matter}}} \textbf{\bibinfo{volume}{9}},
  \bibinfo{pages}{135--139}, \doiprefix\url{10.1007/s10035-006-0019-x}
  (\bibinfo{year}{2006}).

\bibitem{Barrat2002}
\bibinfo{author}{Barrat, A.} \& \bibinfo{author}{Trizac, E.}
\newblock \bibinfo{journal}{\bibinfo{title}{Molecular dynamics simulations of
  vibrated granular gases}}.
\newblock {\emph{\JournalTitle{Physical Review E}}}
  \textbf{\bibinfo{volume}{66}}, \bibinfo{pages}{051303},
  \doiprefix\url{10.1103/PhysRevE.66.051303} (\bibinfo{year}{2002}).

\bibitem{Meer2006}
\bibinfo{author}{van~der Meer, D.} \& \bibinfo{author}{Reimann, P.}
\newblock \bibinfo{journal}{\bibinfo{title}{Temperature anisotropy in a driven
  granular gas}}.
\newblock {\emph{\JournalTitle{Europhysics Letters ({EPL})}}}
  \textbf{\bibinfo{volume}{74}}, \bibinfo{pages}{384--390},
  \doiprefix\url{10.1209/epl/i2005-10552-9} (\bibinfo{year}{2006}).

\bibitem{Noije1998}
\bibinfo{author}{van Noije, T.} \& \bibinfo{author}{Ernst, M.}
\newblock \bibinfo{journal}{\bibinfo{title}{Velocity distributions in
  homogeneous granular fluids: the free and the heated case}}.
\newblock {\emph{\JournalTitle{Granular Matter}}} \textbf{\bibinfo{volume}{1}},
  \bibinfo{pages}{57--64}, \doiprefix\url{10.1007/s100350050009}
  (\bibinfo{year}{1998}).

\bibitem{Binder1987}
\bibinfo{author}{Binder, K.}
\newblock \bibinfo{journal}{\bibinfo{title}{Theory of first-order phase
  transitions}}.
\newblock {\emph{\JournalTitle{Reports on Progress in Physics}}}
  \textbf{\bibinfo{volume}{50}}, \bibinfo{pages}{783--859},
  \doiprefix\url{10.1088/0034-4885/50/7/001} (\bibinfo{year}{1987}).

\bibitem{Weber1995}
\bibinfo{author}{Weber, H.}, \bibinfo{author}{Marx, D.} \&
  \bibinfo{author}{Binder, K.}
\newblock \bibinfo{journal}{\bibinfo{title}{Melting transition in two
  dimensions: A finite-size scaling analysis of bond-orientational order in
  hard disks}}.
\newblock {\emph{\JournalTitle{Phys. Rev. B}}} \textbf{\bibinfo{volume}{51}},
  \bibinfo{pages}{14636--14651}, \doiprefix\url{10.1103/PhysRevB.51.14636}
  (\bibinfo{year}{1995}).

\bibitem{Zahn2000}
\bibinfo{author}{Zahn, K.} \& \bibinfo{author}{Maret, G.}
\newblock \bibinfo{journal}{\bibinfo{title}{Dynamic criteria for melting in two
  dimensions}}.
\newblock {\emph{\JournalTitle{Physical Review Letters}}}
  \textbf{\bibinfo{volume}{85}}, \bibinfo{pages}{3656},
  \doiprefix\url{10.1103/PhysRevLett.85.3656} (\bibinfo{year}{2000}).

\bibitem{McNamara1992}
\bibinfo{author}{McNamara, S.} \& \bibinfo{author}{Young, W.}
\newblock \bibinfo{journal}{\bibinfo{title}{Inelastic collapse and clumping in
  a one‐dimensional granular medium}}.
\newblock {\emph{\JournalTitle{Physics of Fluids A: Fluid Dynamics}}}
  \textbf{\bibinfo{volume}{4}}, \bibinfo{pages}{496--504},
  \doiprefix\url{10.1063/1.858323} (\bibinfo{year}{1992}).

\bibitem{Xing2021}
\bibinfo{author}{Xing, Y.} \emph{et~al.}
\newblock \bibinfo{journal}{\bibinfo{title}{X-ray tomography investigation of
  cyclically sheared granular materials}}.
\newblock {\emph{\JournalTitle{Phys. Rev. Lett.}}}
  \textbf{\bibinfo{volume}{126}}, \bibinfo{pages}{048002},
  \doiprefix\url{10.1103/PhysRevLett.126.048002} (\bibinfo{year}{2021}).

\bibitem{Andreotti2013}
\bibinfo{author}{Andreotti, B.}, \bibinfo{author}{Forterre, Y.} \&
  \bibinfo{author}{Pouliquen, O.}
\newblock \emph{\bibinfo{title}{Granular Media: Between Fluid and Solid}}
  (\bibinfo{publisher}{Cambridge University Press},
  \bibinfo{address}{Cambridge}, \bibinfo{year}{2013}).

\bibitem{Blair2008}
\bibinfo{author}{Blair, D.} \& \bibinfo{author}{Dufrense, E.}
\newblock \bibinfo{title}{The {M}atlab particle tracking code repository}
  (\bibinfo{year}{2008}).

\bibitem{Rycroft2006}
\bibinfo{author}{Rycroft, C.~H.}, \bibinfo{author}{Grest, G.~S.},
  \bibinfo{author}{Landry, J.~W.} \& \bibinfo{author}{Bazant, M.~Z.}
\newblock \bibinfo{journal}{\bibinfo{title}{Analysis of granular flow in a
  pebble-bed nuclear reactor}}.
\newblock {\emph{\JournalTitle{Phys. Rev. E}}} \textbf{\bibinfo{volume}{74}},
  \bibinfo{pages}{021306}, \doiprefix\url{10.1103/physreve.74.021306}
  (\bibinfo{year}{2006}).

\end{thebibliography}
\end{document}